\documentclass[11pt]{article}
\usepackage{amssymb,amsmath,amsfonts}
\usepackage{graphicx}
\usepackage{graphics}
\usepackage{eepic,epsfig}
\usepackage{verbatim}
\usepackage{color}
\usepackage{hyperref}
\usepackage{cite}

1.60cm \makeatletter \@addtoreset{equation}{section}

\makeatother

\begin{document}
\title{Induced Fermionic vacuum polarization in dS spacetime with a compactified cosmic string}
\author{E. A. F. Bragan\c{c}a$^1$\thanks{E-mail: braganca@df.ufpe.br},  E. R. Bezerra de Mello $^2$\thanks{E-mail: emello@fisica.ufpb.br} \ and A. Mohammadi$^1$\thanks{E-mail: azadeh.mohammadi@df.ufpe.br} \,
 \\\
\textit{$^{1}$Departamento de F\'{i}sica, Universidade Federal de Pernambuco, 52171-900, Recife, PE, Brazil}\\
\textit{$^{2}$Departamento de F\'{\i}sica, Universidade Federal da Para\'{\i}%
ba 58.059-970, Caixa Postal 5.008,}\\
\textit{Jo\~{a}o Pessoa, PB, Brazil}\vspace{%
0.3cm}\\}
\maketitle
%
\begin{abstract}
We study the fermionic condensate (FC) and the vacuum expectation value (VEV) of the energy-momentum tensor
for a massive spinor field in the de Sitter (dS) spacetime including an ideal cosmic string. In addition, 
spatial dimension along the string is compactified to a circle of length $L$. The fermionic field is assumed to
obey quasi-periodic condition along the $z$-axis. There are also magnetic fluxes running along the cosmic string and enclosed by the compact dimension. Both, the FC and the VEV of the energy-momentum tensor,
are decomposed into two parts: one induced by the cosmic string in dS spacetime considering the absence of
the compactification, and another one induced by the compactification. In particular, we show that the FC vanishes for a massless fermionic field.
\end{abstract}
%
%

\section{Introduction}
\label{Int}
It is well known that in the early Universe different types of topological defects may have been formed due to the series of phase transitions \cite{VS} which
among them, cosmic strings have been extensively studied in the literature.
Observations of the cosmic microwave background have ruled cosmic strings out as the main source of the primordial density fluctuations, but several other interesting physical effects can be associated with this topological defect like emission of gravitational waves, generation
of high-energy cosmic rays and doubling images of distant objects \cite{damour2000,bhattacharjee,berezinsky2001}.
Furthermore, a variant in the mechanism of formation for cosmic strings has been proposed in the context of brane inflation
\cite{sarangi2002,copeland2004cosmic,dvali2004formation} leading to a renewed interest in the research on this object. The topological defects can also form in condensed matter systems due to symmetry breaking phase transitions. The conical space can appear as an effective background geometry in certain condensed matter systems such as nanotubes, superfluids, superconductors, crystals, liquid crystals and quantum liquids \cite{nelson2002defects,volovik2003universe}.

The geometry of the spacetime associated with an infinitely long and straight cosmic string is characterized by a planar angle deficit
on the two-surface orthogonal to the string. Besides that, the spacetime is locally flat except on the top of the string where
a delta shaped curvature tensor is present. Cosmic strings were first introduced in the literature as being created by a Dirac-delta type
distribution of energy and an axial stress, however it can also be described in the context of a classical field theory which the energy-momentum tensor associated with the vortex configuration
of the Maxwell-Higgs system \cite{Nielsen197345} couples to the Einstein equations. Garfinkle \cite{PhysRevD.32.1323} and Linet \cite{Linet1987240} have studied this coupled system and have shown that a planar angle deficit arises on the
two-surface perpendicular to the string, as well as a magnetic flux running along to its core. One of the most remarkable features of this spacetime is the fact that fields are sensitive to its global conical structure which can cause interesting phenomena.

It is well known that geometry and topology play important roles in many physical problems with implications from subnuclear to cosmological scales. 
In the context of the quantum field theory, the vacuum properties
are influenced by both geometrical and topological aspects of the background spacetime. In the present paper, we intend to investigate
the combined effects of the geometry and topology on the fermionic condensate and the VEV of the energy-momentum tensor associated
with a massive fermion field. We take the de Sitter spacetime as the background geometry and that the topological effects
are induced by the presence of a cosmic string and the compactification of the spatial dimension along the string.

The dS spacetime is the curved spacetime most analyzed in the context of quantum field theory and cosmology. Considering the presence of a positive cosmological constant, the dS spacetime is a maximally symmetric solution
of Einstein equations. As a consequence of this high degree of symmetry, several numbers of physical problems can
be exactly solved in this spacetime. In addition, the importance of this background has increased after the advent
of the Universe's expansion in its early stages from an inflationary scenario.  In this scenario the geometry of the Universe can be
approximated by a portion of the dS spacetime and due to this fact some problems in the standard cosmology are naturally solved.
Besides that, during the inflationary epoch, fluctuations in the inflaton field have generated inhomogeneities that play an important
role in the generation of large scale structures in the Universe \cite{linde2008inflationary}. 

The type of topological effect we shall consider here is induced by a planar angle deficit due the presence of a cosmic string. The conical structure of the spacetime associated with this topological defect modifies the vacuum
fluctuations associated to quantum fields. In this way, the vacuum expectation values (VEVs) of the energy-momentum tensor associated with
scalar \cite{PhysRevD.35.536,escidoc:153364,GL,DS,PhysRevD.35.3779}
and fermion \cite{LB,Moreira1995365,BK} fields present a nonzero value.
Additional contributions to the above VEVs associated with charged quantum fields
are induced when is considered the presence of a magnetic flux running along the string
\cite{PhysRevD.36.3742,guim1994,SBM,SBM2,SBM3}. Vacuum current densities are also induced by the magnetic fluxes and it has been shown that the azimuthal induced current density arises if the ratio of the magnetic flux by the quantum one has a nonzero fractional part \cite{LS,SNDV}. In addition, induced current densities in a higher-dimensional cosmic string spacetime \cite{ERBM} and in a $(2+1)$-dimensional cosmic string spacetime with a circular boundary \cite{PhysRevD.82.085033}
in the presence of a magnetic flux were also studied.
In the aforementioned analysis the cosmic string had no inner structure, i.e., it is considered as an ideal linear object. The inner structure of the cosmic string has been taken into consideration in Refs. \cite{deMello:2014ksa,de2016induced} for the scalar and fermionic currents and also for the energy-momentum tensor in
\cite{de2017fermionic}.

The compactification along the cosmic string axis can also induce a topological effect in the system. Most high-energy theories of fundamental physics, like supergravity and superstring theories, have this feature in common. An interesting application of a theoretical model with a compact dimension appeared in the context of nanophysics.
In cylindrical and toroidal carbon nanotubes, the
background of the corresponding field theory presents compact dimensions \cite{RevModPhys.81.109}. In the context of quantum
field theory, the imposition of periodicity conditions on the field operator along compact dimension also alters the VEVs of physical observables. In \cite{PhysRevD.82.065011} and \cite{bellucci2013fermionic}
the fermionic current in the presence of an arbitrary number of compactified spatial dimensions and a constant gauge was investigated. The VEV of the induced fermionic current and the energy-momentum tensor considering
a compactified cosmic string spacetime in the presence of a magnetic flux running through the string was studied in \cite{BMSAA,SERA}. Moreover, for a charged scalar field, the VEVs of the induced bosonic current and energy-momentum tensor in a higher dimensional compactified cosmic string spacetime was considered in \cite{Braganca:2014qma} and \cite{Braganca:2019mvj}, respectively. 
For Schwarzschild spacetime equipped by a cosmic string, the vacuum polarization was investigated in \cite{ottewill2010vacuum} and \cite{ottewill2010renormalized}.
For the dS spacetime, the vacuum polarization induced by a cosmic string for a scalar field \cite{BezerradeMello:2009ng} and a fermion field \cite{BezerradeMello:2010ci}, and the calculation of the vacuum fermionic current \cite{mohammadi2015induced} have been developed. In addition, similar analysis induced by a cosmic string
in anti-dS spacetime  have also been considered for massive scalar \cite{de2012vacuum} and fermion fields \cite{de2013fermionic}, and for a scalar field with a compactified extra dimension \cite{Santos:2018ttf}. With the intention of developing a further analysis, here we plan to investigate the fermionic vacuum polarization and the VEV of the energy-momentum tensor in dS spacetime, considering the presence of a compactified cosmic string. In this way, this present analysis is as general as possible. However, in the limit where the size of the compactification goes to infinity one recovers the result in the absence of the compactification.


The paper is organized as follows. In Section \ref{fmodes} we describe the background geometry and the complete set of normalized positive- and negative-energy fermionic mode function which obey a quasiperiodic boundary condition with an arbitrary phase along the $z$-axis. We also assume the presence of a constant gauge field.
In Section \ref{FC}, by using the mode-summation method we develop the analysis of the FC which is decomposed into two contributions: the first one corresponds to the geometry of a cosmic string in dS spacetime with no compactification and the second one induced by the compactification of the spatial dimension along the string. In Section \ref{EMTsec} we develop the calculation of the VEV of the energy-momentum tensor,
proving the same decomposition. In section \ref{EMTsec2} we study some properties of the results obtained in the previous section and discuss some limiting cases.
Our conclusions are summarized in Section \ref{conc}. Throughout the paper we use natural units in which $G=\hbar=c=1$.

\section{Background geometry and fermionic modes}
\label{fmodes}
The line element describing a cosmic string along the $z$-axis in dS spacetime is given by
\begin{equation}
ds^2=g_{\mu\nu}dx^\mu dx^\nu=dt^2- e^{2t/\alpha}\left(dr^2+r^2 d\phi^2+dz^2\right),
\label{line}
\end{equation}
in cylindrical coordinates with non-negative $r$, $t\in(-\infty,+\infty)$ and $\phi\in[0,2\pi/q]$ where the presence of the cosmic
string is codified by the parameter $q>1$. The parameter $\alpha$ appearing
in the line element above is related to the Ricci scalar $R$ and the cosmological constante $\Lambda$ as  $R=12\alpha^{-2}$ and $\Lambda=3\alpha^{-2}$.
Making use of the conformal time $\eta$, which is defined as \cite{BezerradeMello:2009ng}
\begin{equation}
\eta=\alpha e^{-t/\alpha} \ \ \ \ , \ \ \ \ \eta\in[0,\infty],
\end{equation}
we can write the line element \eqref{line} in the following form
\begin{equation}
ds^2=(\alpha/\eta)^2\left(d\eta^2-dr^2-r^2 d\phi^2-dz^2\right).
\end{equation}
Besides that, the direction along the $z$-axis is compactified to a circle with length $L$, meaning $z\in[0,L]$, and we assume
that along this compact direction the fermionic field obeys the following quasiperiodicity condition
\begin{equation}
\psi (t,r,\phi,z+L)=e^{2\pi i \beta}\psi (t,r,\phi,z).
\label{quasiperiodic}
\end{equation}
The parameter $\beta$ in the above expression is defined as a constant in the interval $[0,1]$. We have two special cases for this parameter: for
$\beta=0$ we have a peridioc boundary condition, while for $\beta=1/2$ we have an antiperiodic boundary condition. These values correspond to
untwisted and twisted fields, respectively.

The dynamics of a massive spinor field in the above curved spacetime with a magnetic flux running along the string is governed by the Dirac equation in the following form
\begin{equation}
i\gamma^\mu \mathcal{D}_\mu\psi -m\psi=0, \ \ \ \mathcal{D}_\mu=\partial_\mu +ieA_\mu+\Gamma_\mu,
\label{diracequation}
\end{equation}
knowing that $\Gamma_\mu$ and $A_\mu$ are the spin connection and a four vector potential, respectively. 
Note that the physical components of the vector potential, $A_\phi$ and $A_z$,
are related to the covariant components by $A_\phi=-A_2/r$ and $A_z=-A_3$ in the Dirac equation \eqref{diracequation}.
To obtain the fermionic condensate as well as the vacuum expectation value of the energy-momentum tensor one needs a complete set of fermionic modes. The positive- and negative-energy fermionic modes in this spacetime background were obtained in \cite{mohammadi2015induced} which are
\begin{equation}
\psi_\sigma^{(\pm)}(x)=c^{(\pm)}_\sigma\eta^2 e^{iq(j+a)\phi+ikz}
\left(
\begin{array}{c}
H^{(\lambda_{\pm})}_{1/2-im\alpha}(\gamma\eta)J_{\beta_1}(pr)e^{-iq\phi/2}\\
\\
\frac{i\, s\,p\, \epsilon_j}{\gamma+sk}H^{(\lambda_{\pm})}_{1/2-im\alpha}(\gamma\eta)J_{\beta_2}(pr)e^{iq\phi/2}\\
\\
-i\, s\, H^{(\lambda_{\pm})}_{-1/2-im\alpha}(\gamma\eta)J_{\beta_1}(pr)e^{-iq\phi/2}\\
\\
\frac{p\,\epsilon_j}{\gamma+sk}H^{(\lambda_{\pm})}_{-1/2-im\alpha}(\gamma\eta)J_{\beta_2}(pr)e^{iq\phi/2}
\end{array}%
\right) ,
\label{modes}
\end{equation}
where the gauge transformation $A_\mu \to A_\mu+\partial_\mu \Lambda$, $\psi \to e^{-ie\Lambda}\psi$ with $\Lambda=-A_\mu x^{\mu}$ leading to the new Dirac equation, $(i\gamma^\mu \nabla_\mu-m)\psi=0$, was used. The above fermionic modes are
specified by the set $\sigma =(p,k,j,s)$ of quantum numbers.
In the above expression, $\lambda_+=1$, $\lambda_-=2$, $j=\pm 1/2,\pm 3/2, ...$, and $s=\pm 1$. In addition, $J_\nu(x)$ and $H^{(1,2)}_\nu(x)$ are Bessel and Hankel functions \cite{abramowitz}, respectively, and $\gamma=\sqrt{k^2+p^2}$ where $0\leq p<\infty$. Moreover, in the Eq. \eqref{modes} we have
\begin{equation}
\beta_1=q|j+a|-\epsilon_j/2, \quad \beta_2=q|j+a|+\epsilon_j/2, \quad |c_\sigma^{(\pm)}|^2=\frac{qpe^{\pm m\alpha\pi}}{16L\alpha^3}(\gamma+sk),
\label{aparam}
\end{equation}
with $k=k_l=2\pi(l+\tilde{\beta})/L$, $l=0, \pm 1,\pm 2,...$ and
\begin{equation}
a=\frac{\Phi_\phi}{\Phi_0} \ \ \ {\rm and} \ \ \ \tilde{\beta}=\beta-\frac{\Phi_z}{\Phi_0},
\end{equation}
where $\Phi_0=2\pi/e$ is the quantum flux while $\Phi_\varphi$ and $\Phi_z$ are the magnetic fluxes in the azimuthal and axial directions,
respectively. Besides that, $\epsilon_j=1$ for $j>-a$ and $\epsilon_j=-1$ for $j<-a$.

One can write the parameter $a$ in the form
\begin{equation}
a=n_0+a_0, \quad |a_0|<1/2,
\label{a0}
\end{equation}
with $n_0$ being an integer number. It is easy to see that the VEVs of the physical observables are given in terms of $a_0$ solely, by simply a shift like $j+n_0 \to j$.

In \cite{mohammadi2015induced}, we studied the vacuum fermionic currents in the same geometry, a compactified cosmic string in the background of dS spacetime assuming that the field is prepared in the Bunch-Davies vacuum state. In the continuation of the work, we investigate the fermionic condensate and the renormalized vacuum expectation value of the energy-momentum tensor in the present paper.

\section{Fermionic condensate}
\label{FC}
In this section we develop the analysis of the FC, which is defined by the VEV
\begin{equation}
\langle 0|\bar{\psi}\psi |0\rangle \equiv \langle \bar{\psi}\psi \rangle=\sum_{\sigma}\bar{\psi}^{(-)}_\sigma \psi_\sigma^{(-)}.
\end{equation}
with $|0\rangle$ representing the vacuum state and $\bar{\psi}=\psi^\dagger\gamma^0$ being the Dirac adjoint.
By using the mode functions \eqref{modes} one can write the FC as
\begin{eqnarray}
\langle \bar{\psi}\psi \rangle&=&\frac{q\eta^4e^{-m\alpha\pi}}{16L\alpha^3}\sum_\sigma p\, (\gamma+sk)
\left[J^2_{\beta_1}(pr)+\frac{p^2}{(\gamma+sk)^2}J^2_{\beta_2}(pr)\right]\nonumber\\
&& \times \left[|H^{(2)}_{1/2-im\alpha}(\gamma\eta)|^2
-|H^{(2)}_{-1/2-im\alpha}(\gamma\eta)|^2\right],
\end{eqnarray}
using the compact notation for the sum over all independent quantum numbers as
\begin{equation}
\sum_\sigma=\int_0^\infty dp \, \sum_{l=-\infty}^{+\infty}\, \sum_{s=\pm1}\, \sum_{j=\pm 1/2, \pm 3/2,...}.
\end{equation}
Making use of \eqref{App} and evaluating the summation over $s$ result in
\begin{eqnarray}
\langle \bar{\psi}\psi \rangle&=&\frac{q\eta^4}{2\pi^2L\alpha^3}\int_0^\infty dp \sum_{l=-\infty}^{+\infty}\sum_{j}
\gamma\, p\left[J^2_{\beta_1}(pr)+J^2_{\beta_2}(pr)\right]\nonumber\\
&\times&\left[|K_{1/2-im\alpha}(i\gamma\eta)|^2-|K_{1/2+im\alpha}(i\gamma\eta)|^2\right].
\end{eqnarray}
By means of the relation \eqref{App2} we obtain
\begin{eqnarray}
\langle \bar{\psi}\psi \rangle&=&-\frac{iq\eta^4}{2\pi^2\alpha^3 L}\left(\partial_\eta+\frac{1-2im\alpha}{\eta}\right)
\int_0^\infty dp \, p\sum_j\left[J^2_{\beta_1}(pr)+J^2_{\beta_2}(pr)\right]\nonumber\\
&&\times\sum_{l=-\infty}^{\infty}K_{1/2-im\alpha}(i\gamma\eta)K_{1/2-im\alpha}(-i\gamma\eta) \  ,
\end{eqnarray}
being $\gamma=\gamma_l=\sqrt{p^2+(2\pi(l+\tilde{\beta})/L)^2}$ and  $K_\nu(z)$ the Macdonald function \cite{abramowitz}. The summation over $l$ can be evaluated with the help of the Abel-Plana formula in the form
\cite{Saharian:2007ph,PhysRevD.78.045021,PhysRevD.79.085019}
\begin{eqnarray}
\frac{2\pi}{L}\sum_{l=-\infty}^{+\infty}g(k_l)f(|k_l|)&=&\int_0^\infty du\, [g(u)+g(-u)]f(u)+i\int_0^{\infty}
du\, [f(iu)-f(-iu)]\nonumber\\
&&\times\sum_{\chi=\pm1}\frac{g(i\chi u)}{e^{Lu+2\pi i\chi\tilde{\beta}}-1},
\label{AP}
\end{eqnarray}
where one can decompose the FC in the following way
\begin{equation}
\langle \bar{\psi}\psi \rangle=\langle \bar{\psi}\psi \rangle_s^{dS}+\langle \bar{\psi}\psi \rangle_c.
\end{equation}
The first term in the right-hand side of the above equation is the contribution for the FC due to the curvature of the dS spacetime in the cosmic string background with no compactification while the second one is the contribution induced by the compactification along the z-axis. Obviously, the second term should vanish in the limit $L\rightarrow\infty$.

By choosing $g(u)=1$ and
\begin{equation}
f(u)=K_{1/2-im\alpha}(i\gamma\eta)K_{1/2-im\alpha}(-i\gamma\eta),
\label{ufunction}
\end{equation}
the first term in the Eq. \eqref{AP} gives
\begin{eqnarray}
\langle \bar{\psi}\psi \rangle_s^{dS}&=&-\frac{iq\eta^4}{2\pi^3\alpha^3}\left(\partial_\eta+\frac{1-2im\alpha}{\eta}\right)\int_0^\infty dp \, p  \sum_{j}\left[J^2_{\beta_1}(pr)+J^2_{\beta_2}(pr)\right]\nonumber\\
&&\times\int_0^{\infty}dk \ K_{1/2-im\alpha}(i\gamma\eta)K_{1/2-im\alpha}(-i\gamma\eta).
\end{eqnarray}

Using the integral representation \eqref{MacRep} as well as the relation \eqref{bess}, the integral over $k$ and $p$ can be done directly which gives
\begin{eqnarray}
\langle \bar{\psi}\psi\rangle_s^{dS}&=&-\frac{iq}{2^{9/2}\pi^{5/2}\alpha^3}\left(\eta\frac{\partial}{\partial\eta}+1-2im\alpha\right)
\int_0^\infty dy\ \frac{\cosh[(1-2im\alpha)y]}{\sinh^3y}\nonumber\\
&&\times\int_0^\infty \frac{du}{u^{5/2}}\   e^{-1/(2u)-r^2/(4u\eta^2\sinh^2y)}
\ \mathcal{J}\left(q,a_0,\frac{r^2}{4u\eta^2\sinh^2y}\right) ,\nonumber\\
\end{eqnarray}
with the notation
\begin{equation}
\mathcal{J}(q,a_0,w)=\sum_j[I_{\beta_1}(w)+I_{\beta_2}(w)].
\label{summationj1}
\end{equation}
Defining a new variable $x\equiv r^2/(4u\eta^2\sinh^2y)$ the integral over $y$ can be evaluated directly
and we obtain
\begin{eqnarray}
\langle \bar{\psi}\psi\rangle_s^{dS}&=&-\frac{iq\eta^3}{(2\pi)^{5/2}(\alpha r)^3}\left(\eta\frac{\partial}{\partial\eta}+1-2im\alpha\right)\int_0^\infty dx\,x^{1/2}e^{x(\eta^2/r^2-1)}\nonumber\\
&&\times K_{1/2-im\alpha}(x\eta^2/r^2)\ \mathcal{J}(q,a_0,x),
\end{eqnarray}
Now, considering the formula \eqref{Macder} and defining a new variable $z\equiv x\eta^2/r^2$, we have
\begin{align} \label{FCS1}
\langle \bar{\psi}\psi\rangle_s^{dS}=\frac{q}{\sqrt{2}\pi^{5/2}\alpha^3}\int_0^\infty dz \,z^{3/2}e^{z(1-r^2/\eta^2)} \, {\rm{Im}}\, [K_{1/2-im\alpha}(z)]
\mathcal{J}(q,a_0,zr^2/\eta^2),
\end{align}
Following the procedure used in the Ref. \cite{Bellucci:2011zr} for the summation over $j$ in \eqref{summationj1},
we obtain
\begin{eqnarray}
\mathcal{J}(q,a_0,w)&=&\frac{2}{q}e^{w}+\frac{4}{q}\sum_{k=1}^p(-1)^k\cos(2k\pi a_0)\cos(k\pi/q)e^{w\cos(2k\pi/q)}\nonumber\\
&&+\frac{4}{\pi}\int_0^\infty dx \frac{h(q,a_0,2x)\sinh x}{\cosh(2qx)-\cos(q\pi)}\, e^{-w\cosh 2x},
\label{representation}
\end{eqnarray}
where $p$ is an integer number defined by the relation $2p<q<2p+2$ and
\begin{equation}
h(q,a_0,2x)=\sum_{\chi=+,-}\cos[q\pi(1/2+\chi a_0)]\sinh[2qx(1/2-\chi a_0)].
\end{equation}
In the case of $q<2$ the summation term in \eqref{representation} is absent. Also for the special case of $q=2p$,
one should add the term
\begin{equation}
-(-1)^{q/2}\frac{e^{-w}}{q}\sin(q\pi a_0)
\end{equation}
to the Eq. \eqref{representation}.

The first term in the right-hand side of the Eq. \eqref{representation} provides the FC due only to the dS spacetime. The second and the third ones are the contributions due to the non-trivial topology of the cosmic string spacetime and the magnetic flux. Because here we are mostly interested in the contribution to the FC induced by the presence of the string, we shall discard the pure dS contribution, $\langle \bar{\psi}\psi\rangle_{dS}$, according to the expression below,
\begin{equation}
\langle \bar{\psi}\psi\rangle_s=\langle \bar{\psi}\psi\rangle_s^{dS}-\langle \bar{\psi}\psi\rangle_{dS}  \  .
\label{nondiv}
\end{equation}
It is important to note that the presence of cosmic string does not change the local geometry of dS spacetime for points away from the string. Consequently, the renormalization is needed only for the pure dS contribution. Therefore, the result is finite  for $r\neq 0$ and is given by 
\begin{align}
\langle \bar{\psi}\psi\rangle_s&=\frac{2^{3/2}}{\pi^{5/2}\alpha^3}
\left[\sum_{k=1}^p(-1)^k\cos(2k\pi a_0)\cos(k\pi/q)\int_0^\infty dz \, z^{3/2}e^{z[1-2(r^2/\eta^2)s^2_k]}\right.\nonumber\\
&\left.+\, \frac{q}{\pi}\int_0^\infty dx\frac{h(q,a_0,2x)\sinh x}{\cosh(2qx)-\cos(q\pi)}
\int_0^\infty dz \, z^{3/2}e^{z[1-2(r^2/\eta^2)c^2_x]}\right]
{\rm{Im}}[K_{1/2-im\alpha}(z)],
\label{FCS2}
\end{align}
where we have defined
\begin{equation}
s_k=\sin(k\pi/q) \ \ {\rm and} \ \ c_x=\cosh x.
\end{equation}
Using the formula \eqref{leg1} we can 
evaluate the integral over $z$ and also with the help of \eqref{leg2} the final expression for the FC induced by the
magnetic flux and the planar angle deficit reads
\begin{align}
\langle \bar{\psi}\psi\rangle_s&=\frac{2^{3/2}}{\pi^{5/2}\alpha^3}
\left\{\sum_{k=1}^p(-1)^k\cos(2k\pi a_0)\cos(k\pi/q) \, \mathcal{G}_0 (u_{0,k})\right.\nonumber\\
&\left.+\frac{q}{\pi}\int_0^\infty dx\frac{h(q,a_0,2x)\sinh x}{\cosh(2qx)-\cos(q\pi)} \, \mathcal{G}_0 (u_{0,x})\right\}.
\label{FCS3}
\end{align}
In the above expression we have introduced the notation
\begin{align}
\mathcal{G}_l(u_{l,y})={\rm Im}\left[\mathcal{F}_l(u_{l,y})\right],
\label{Gfunction}
\end{align}
where
\begin{align}
\mathcal{F}_l(u_{l,y})=\frac{\sqrt{\pi}}{2^{7/2}}\, \frac{\Gamma(3-i m \alpha)\Gamma(2+im\alpha)}
{u_{l,y}^{3-im\alpha}}\, F(a,b;c;d_{l,y}),
\label{Gfunction2}
\end{align}
with $F(a,b;c;d_{l,y})$ being the hypergeometric function
\begin{equation}
F(a,b;c;d_{l,y}) \equiv F\left(2-\frac{im\alpha}{2},\frac{3-im\alpha}{2};3;1-u_{l,y}^{-2}\right)
\label{udefinition}
\end{equation}
 and
\begin{equation}
u_{l,k}=\frac{l^2L^2}{2\eta^2}+\frac{2r^2s_k^2}{\eta^2}-1 \ \ {\rm and} \ \
u_{l,x}=\frac{l^2L^2}{2\eta^2}+\frac{2r^2c_x^2}{\eta^2}-1.
\label{udefinition}
\end{equation}
In the Eq. \eqref{FCS3}, we have \eqref{Gfunction} and \eqref{udefinition} with $l=0$. One can notice that the string contribution in FC is an even function of the parameter $a_0$.

Now, let us consider some particular cases of the expression \eqref{FCS3}. For example, for a massless fermionic field, it is easy to see that the string part of the FC vanishes. This result is more evident looking at the expression \eqref{FCS2}.

Now, let us study the short and large distances from the string. For the region near the string, $r/\eta\ll 1$, it is more convenient to consider the expression \eqref{FCS2}. In this regime the dominant contribution to this expression comes from large values of $z$ and we can use the expansion of the Macdonald function for large arguments \cite{abramowitz}. The leading term is given by
\begin{align}
\langle \bar{\psi}\psi\rangle_s&\approx-\frac{m}{2\pi^2}\left(\frac{\eta}{\alpha r}\right)^2
\left\{\sum_{k=1}^p(-1)^k\frac{\cos(2k\pi a_0)\cos(k\pi/q)}{s_k^2}
+\frac{q}{\pi}\int_0^\infty dx\, \frac{c_x^{-2} \, h(q,a_0,2x)\tanh x}{\cosh(2qx)-\cos(q\pi)}\right\}.\nonumber\\
\label{FCR0}
\end{align}
 It is important to have in mind that the ratio $r/\eta$ means the proper distance from the cosmic string in units of the dS curvature ratio $\alpha$.
In the above expression there is a divergence with the second power of the proper distance matching exactly the behavior obtained in \cite{BezerradeMello:2010ci}.

\begin{figure}[h]
	\centering
	{\includegraphics[width=0.48\textwidth]{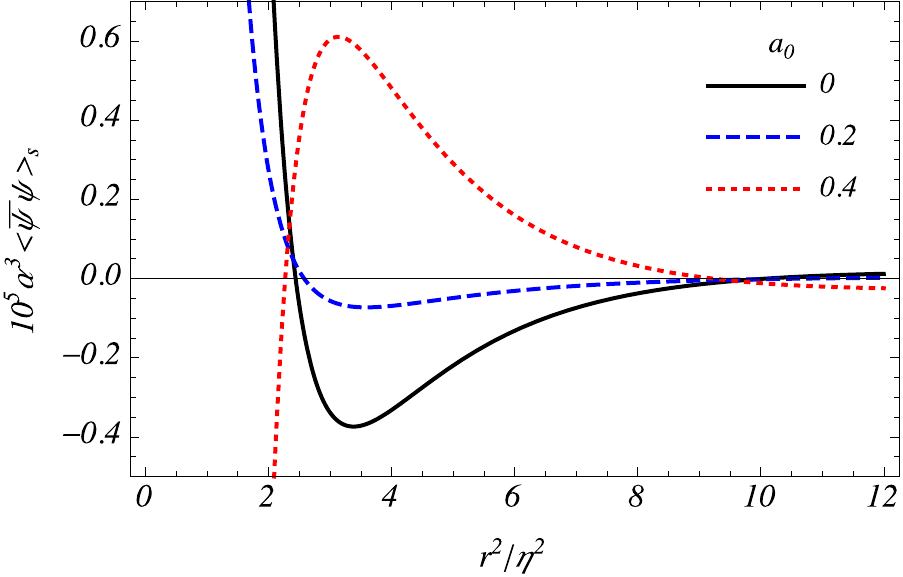}}
	\hfill
	{\includegraphics[width=0.485\textwidth]{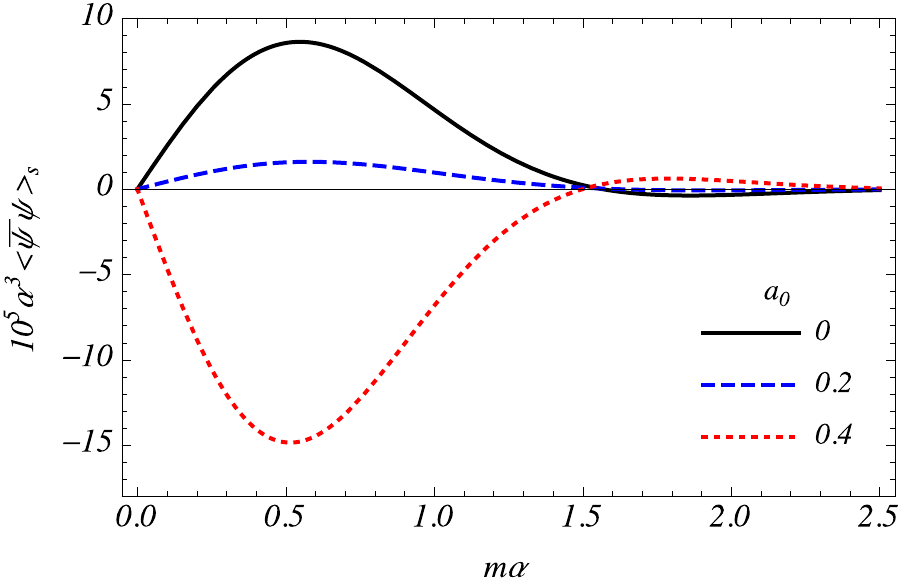}}
	\caption{Fermionic condensate induced by the magnetic flux and planar angle defict as function of $r^2/\eta^2$ (left plot)
	for $m\alpha=2$ and
	the mass of the field (right plot) for $r^2\eta^2=2$. For both plots we adopted $q=1.5$.}
	\label{fig01}
\end{figure}

For large distances from the string, $r/\eta \gg1$, the main contribution to the integral over $z$ in \eqref{FCS2} comes from the
lower limit of integration and we can use the asymptotic expression of the Macdonald function for small arguments. Considering the leading term, the string part of the FC behaves as
\begin{align}
\langle \bar{\psi}\psi\rangle_s&\approx\frac{1}{\pi^{5/2}\alpha^3}\left(\frac{\eta}{r}\right)^4
{\rm Im}\left\{\frac{\Gamma(2+im\alpha)\Gamma(1/2-im\alpha)}{2^{2im\alpha-1}}\left(\frac{\eta}{r}\right)^{im\alpha}
\left[\sum_{k=1}^p(-1)^k\, \frac{\cos(2k\pi a_0)\cos(k\pi/q)}{s_k^{4+2im\alpha}}\right.\right.\nonumber\\
&\times \left. \left. \frac{q}{\pi}\int_0^\infty dx\, \frac{c_x^{-4-2im\alpha} \sinh x\,  h(q,a_0,2x)}{\cosh(2qx)-\cos(q\pi)}\right]\right\}.
\label{FClargeR}
\end{align}
The string part of the FC is exhibited as a function of the ratio $r^2/\eta^2$ in the left plot of Fig. \ref{fig01}, for different values of the parameter $a_0$.\footnote{All the figures presented in this paper are obtained numerically using {\it Mathematica}. }
In the right plot we show the same but now as a function of $m\alpha$, where we note that the FC vanishes in the limit of very massive fields. The results shown in this figure matches the expected result for the case $a_0=0$, in the absence of the magnetic field [see figure 1 in \cite{BezerradeMello:2010ci}].
As one can see, the behavior of the string part of FC as a function of the proper distance and the mass changes dramatically depending on the value of $a_0$. With the parameters chosen for the figure, the threshold value is around $a_0\approx0.22$. 

Now, for the contribution of the FC induced by the compactification using the second term of the Abel-Plana summation formula
we obtain
\begin{align}
\langle \bar{\psi}\psi\rangle_c&=\frac{q\eta^4}{4\pi^3\alpha^3}
\left(\partial_\eta+\frac{1-2im\alpha}{\eta}\right)\int_0^\infty dp \, p\sum_j[J^2_{\beta_1}(pr)+J^2_{\beta_2}(pr)]  \int_p ^\infty du \, K_{1/2-im\alpha}(\eta\sqrt{u^2-p^2})\nonumber\\
& \times \left[K_{1/2-im\alpha}(e^{i\pi}\eta\sqrt{u^2-p^2})-K_{1/2-im\alpha}(e^{-i\pi}\eta\sqrt{u^2-p^2})\right]\sum_{\chi=\pm1}\frac{1}{e^{Lu+2\pi i\chi\tilde{\beta}}-1},
\end{align}
where we have used  the relation
\begin{eqnarray}
\sqrt{(\pm iu)^2+p^2}=\left\{\begin{array}{rc}
\sqrt{p^2-u^2}, & {\rm if}\quad u<p, \\
\pm i\sqrt{u^2-p^2}, &{\rm if}\quad u>p,
\end{array}\right.
\label{Euler}
\end{eqnarray}
to show that the integral over $u$ in the interval $[0,p]$ vanishes while in the interval $[p,\infty]$ does not.
Equations \eqref{Mac1} and \eqref{Mac2} help to write the part of the FC induced by the compactification as follows
\begin{align}
\langle \bar{\psi}\psi\rangle_c&=-\frac{iq\eta^4}{2\pi^2\alpha^3}\sum_{l=1}^{\infty}\cos(2\pi l\tilde{\beta})
\left(\partial_\eta+\frac{1-2im\alpha}{\eta}\right)\int_0^\infty dp\, p\sum_j[J^2_{\beta_1}(pr)+J^2_{\beta_2}(pr)]\nonumber\\
&\times\int_0^\infty d\lambda \lambda\frac{e^{-lL\sqrt{\lambda^2 +p^2}}}{\sqrt{\lambda^2 +p^2}}
K_{1/2-im\alpha}(\eta\lambda)\left[I_{1/2-im\alpha}(\eta\lambda)+I_{-1/2+im\alpha}(\eta\lambda)\right],
\label{FCc1}
\end{align}
where $\lambda=\sqrt{u^2-p^2}$ and we have used the expansion $\left(e^u-1\right)^{-1}=\sum_{l=1}^{\infty}e^{-lu}$.
For the further transformation of the above expression we apply the identity bellow,
\begin{equation}
\frac{e^{-lL\sqrt{\lambda^2 +p^2}}}{\sqrt{\lambda^2 +p^2}}=\frac{2}{\sqrt{\pi}}\int_0^\infty
ds \ e^{-\left(\lambda^2+p^2\right)s^2-l^2L^2/(4s^2)}.
\label{IntRep}
\end{equation}
Substituting the above integral representation into the Eq. \eqref{FCc1}, we are able to evaluate the integration over
$p$ and $\lambda$. Therefore, for the part induced by the compactification we obtain
\begin{eqnarray}
\langle \bar{\psi}\psi\rangle_c&=&\frac{\sqrt{2}q}{\pi^{5/2}\alpha^3}\sum_{l=1}^\infty \cos(2\pi l\tilde{\beta})\int_0^\infty
dz \, z^{3/2}e^{z[1-r^2/\eta^2-l^2L^2/(2\eta^2)]}\nonumber\\
&\times&{\rm Im}\, [K_{1/2-im\alpha}(z)]\mathcal{J}(q,a_0,zr^2/\eta^2),
\end{eqnarray}
where the function $\mathcal{J}(q,a_0,zr^2/\eta^2)$ is given by the Eq. \eqref{representation}.
As a result, the final expression for the FC induced by the compactification is
given by
\begin{eqnarray}
\langle \bar{\psi}\psi\rangle_c&=&\frac{2^{5/2}}{\pi^{5/2}\alpha^3}\sum_{l=1}^\infty \cos(2\pi l\tilde{\beta})
\left\{\sideset{}{'}\sum_{k=0}^p(-1)^k\cos(2k\pi a_0)\cos(k\pi/q) \, \mathcal{G}_l(u_{l,k})\right.\nonumber\\
&
 +& \left.\, \frac{q}{\pi}\int_0^\infty dx\, \frac{h(q,a_0,2x)\sinh x}{\cosh(2qx)-\cos(q\pi)} \,
\mathcal{G}_l(u_{l,x})\right\} \ , 
\label{FCC}
\end{eqnarray}
with the notation \eqref{Gfunction} and \eqref{udefinition}. It is easy to see that the compactification part as well as the string part of FC is zero for the massless fermion.

The compactification part of the FC can be decomposed as
\begin{equation}
\langle \bar{\psi}\psi\rangle_c=\langle \bar{\psi}\psi\rangle_c^{(0)}+\langle \bar{\psi}\psi\rangle_c^{(q,a_0)}.
\label{FCCdecomp}
\end{equation}
The first term on the right-hand side of the above decomposition is the $k=0$ term of \eqref{FCC} with the coefficient $1/2$. This is a pure topological term dependent only on the compactification and the curvature of the dS spacetime and
independent of the radial coordinate and the magnetic flux. This contribution is given by
\begin{equation}
\langle \bar{\psi}\psi\rangle_c^{(0)}=\frac{2^{3/2}}{\pi^{5/2}\alpha^3}\sum_{l=1}^\infty \cos(2\pi l\tilde{\beta}) \,
\mathcal{G}_l(u_{l,0}).
\label{FCCdecomp1}
\end{equation}
The second term on the right-hand side of \eqref{FCCdecomp} is given by
\begin{align}
\langle \bar{\psi}\psi\rangle_c^{(q,a_0)}&=\frac{2^{5/2}}{\pi^{5/2}\alpha^3}\sum_{l=1}^\infty \cos(2\pi l\tilde{\beta})
\left\{\sum_{k=1}^p(-1)^k\cos(2k\pi a_0)\cos(k\pi/q) \, \mathcal{G}_l(u_{l,k})\right.\nonumber\\
&\left. + \, \frac{q}{\pi}\int_0^\infty dx\frac{h(q,a_0,2x)\sinh x}{\cosh(2qx)-\cos(q\pi)} \,
\mathcal{G}_l(u_{l,x})\right\},
\label{FCC2}
\end{align}
which is the contribution to the FC induced by the compactification and the magnetic flux.

We now evaluate some asymptotic expressions of the previous equation. Considering $r/\eta\ll 1$, near the string, the leading
order reads
\begin{align}
\langle \bar{\psi}\psi\rangle_c^{(q,a_0)}&\approx\frac{q \cos[q\pi(1/2+|a_0|)]\Gamma(1/2-q(1/2-|a_0|))}{\pi^{7/2}\alpha^3}
\left(\frac{r}{\sqrt{2}\eta}\right)^{q(1-2|a_0|)-1}\nonumber\\
&\times\sum_{l=1}^{\infty}\cos(2\pi l\tilde{\beta}) \, {\rm Im}
\left\{\frac{\Gamma(2+im\alpha)\Gamma(3/2-im\alpha+q(1-2|a_0|))}{2^{1+im\alpha}\Gamma(2-im\alpha)
(lL/(\sqrt{2}\eta)^{3-im\alpha+q(1-2|a_0|)})}\right\}.
\label{FCCR0}
\end{align}
The FC is finite on the string for $|a_0|=(1-1/q)/2$, vanishes for $|a_0|<(1-1/q)/2$ and
diverges for $|a_0|>(1-1/q)/2$. Note that in the absence of axial magnetic flux, the FC vanishes on the string's core.

For large values of $L/\eta$, we use  \eqref{hyper1}, and to the leading order we find
\begin{align}
\langle \bar{\psi}\psi\rangle_c^{(q,a_0)}&\approx\frac{2^{5/2}}{\pi^{5/2}\alpha^3}\left(\frac{\eta}{L}\right)^{4}
\sum_{l=1}^{\infty}\cos(2\pi l\tilde{\beta})
\, {\rm Im}\left\{\frac{\Gamma(1/2-im\alpha)\Gamma(2+im\alpha)}{2^{im\alpha+1/2}}
\left(\frac{\eta}{L}\right)^{2im\alpha}\right.\nonumber\\
&\left.\times\left[\sum_{k=1}^p\frac{(-1)^k\cos(2k\pi a_0)\cos(k\pi/q)}
{\left(\frac{l^2}{2}+\frac{r^2s_k^2}{L^2}\right)^{2+im\alpha}}+
\frac{q}{\pi}\int_0^\infty dx\, \frac{h(q,a_0,2x)\sinh x}{\cosh(2qx)-\cos(q\pi)}\right.\right.\nonumber\\
&\left.\left.\times\left(\frac{l^2}{2}+\frac{r^2c_x^2}{L^2}\right)^{-2-im\alpha}\right]\right\}.
\label{FCClargeL}
\end{align}
As can be seen the FC, considering the length of the compact dimension much larger than the curvature of the
dS spacetime, presents a fourth power decay.

\begin{figure}[h]
	\centering
	{\includegraphics[width=0.485\textwidth]{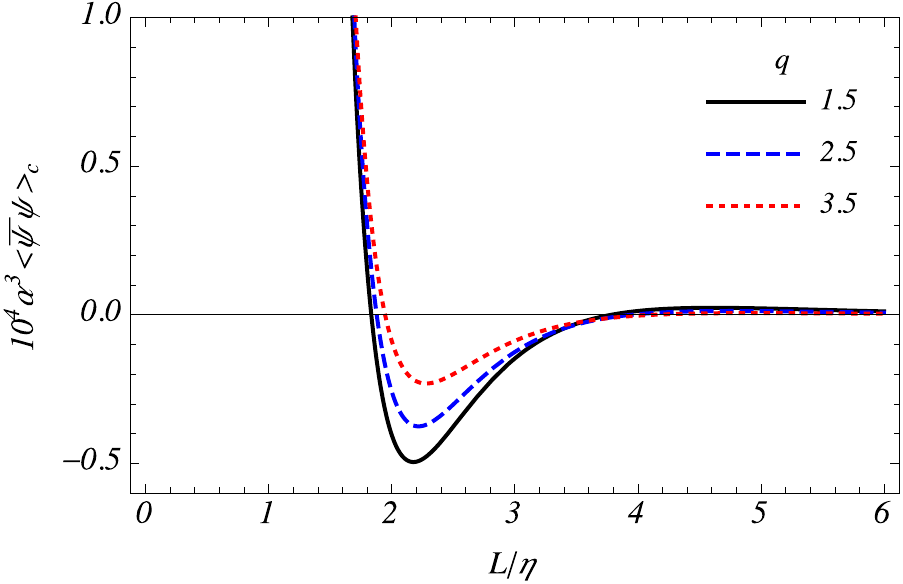}}
	\hfill
	{\includegraphics[width=0.485\textwidth]{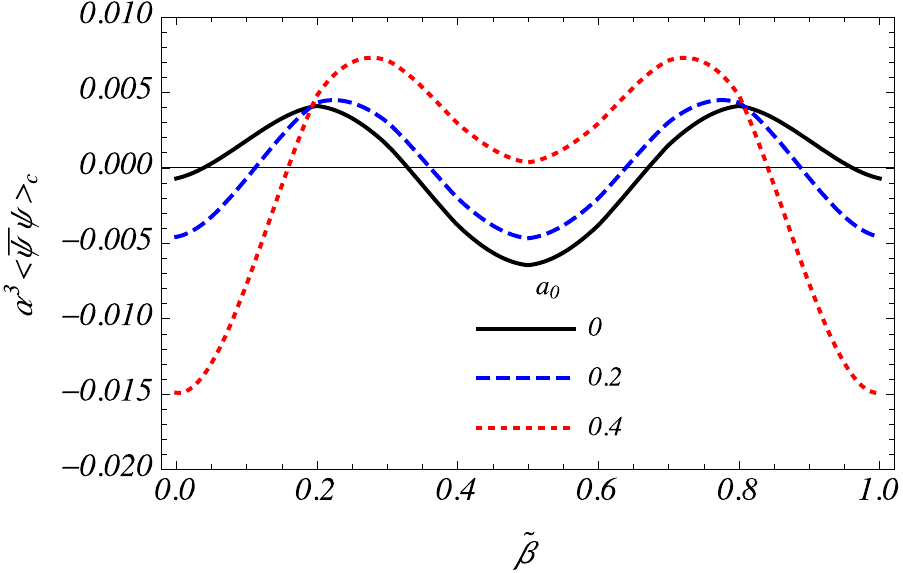}}
	\caption{FC induced by the magnetic flux, the planar angle defict and the compactification. In the left plot the behavior of the FC is shown as a function of
	$L/\eta$ considering three different values of $q$, $a_0=0.25$, $r/\eta=0.75$ and $\tilde{\beta}=0.25$.
	In the right plot we have the same quantity, but as a function of
	$\tilde{\beta}$ considering three different values of $a_0$ for $q=1.5$, $r/\eta=0.5$ and $L/\eta=0.75$. In both graphs  $m\alpha=2$.}
	\label{fig02}
\end{figure}

Figure \ref{fig02} shows the behavior of the FC induced by the compactification as a function of $L/\eta$ (left plot) and $\tilde{\beta}$ (right plot). \footnote{In the summation over $l$, due to the rapid convergence of the series in our numerical calculation, we consider only the dominant contributions given by small $l$. }
As one expects, in the limit $L\rightarrow \infty$, the compactification contribution of FC vanishes.

\section{Energy-momentum tensor}\label{EMTsec}
Another important physical quantity which characterizes the quantum vacuum is the VEV of the energy-momentum tensor.
Considering a charged fermionic field this VEV can be evaluated considering the summation
formula
\begin{equation}
\langle 0| T_{\mu\nu}| 0 \rangle=\frac{i}{2}\sum_\sigma [\bar{\psi}_{\sigma}^{(-)}(x)\gamma_{(\mu}\mathcal{D}_{\nu)}\psi_\sigma^{(-)}(x)
-(\mathcal{D}_{(\mu}\bar{\psi}_\sigma^{(-)}(x))\gamma_{\nu)}\psi_\sigma^{(-)}(x)],
\label{EMT}
\end{equation}
where we consider the negative energy eigenspinors given by the Eq. \eqref{modes} and
$\mathcal{D}_\mu \bar{\psi}=\partial_\mu\bar{\psi}-ieA_\mu\bar{\psi}-\bar{\psi}\Gamma_\mu$. In the above expression,
the brackets in the index expression mean the symmetrization over the enclosed indexes.

The geometry that we have considered allows us to decompose the VEV of the energy-momentum tensor
in the following way
\begin{equation}
\langle 0|T_\mu^\nu|0\rangle=\langle 0|T_\mu^\nu|0\rangle_{s}^{dS}+\langle 0|T_\mu^\nu|0\rangle_c,
\label{decompositionEMT}
\end{equation}
where the first term is the contribution induced by the cosmic string and the curvature of the de Sitter spacetime, while
the second one is the contribution induced by the compactification.

\subsection{Energy density}
Let us start with the evaluation of the vacuum energy density, $\langle 0|T_0^0|0\rangle$. Using the expression for the eigenfunctions given by the Eq. \eqref{modes}, and taking into account that $\Gamma_0=0$ \cite{BezerradeMello:2010ci}, after long
but straightforward calculations, we obtain
\begin{align}
\langle 0| T_0^0| 0 \rangle&=-\frac{q\eta^5}{4\pi^2\alpha^4 L}\sum_{l=-\infty}^{\infty}\sum_j\int_0^\infty dp \, p \,
[J^2_{\beta_1}(pr)+J^2_{\beta_2}(pr)]\nonumber\\
&\times \left[4\gamma^2+\hat{S}_\eta\right]
K_{1/2-im\alpha}(i\gamma\eta)K_{1/2-im\alpha}(-i\gamma\eta),
\label{T00a}
\end{align}
where we have defined the operator
\begin{equation}
\hat{S}_\eta=\partial_\eta^2+\frac{2}{\eta}\partial_\eta+\frac{4im\alpha(1/2-im\alpha)}{\eta^2}.
\end{equation}
The expression for the energy density can be decomposed as indicated by \eqref{decompositionEMT}.
The Abel-Plana summation formula \eqref{AP} allows us to solve the summation over the quantum number $l$ in the
expression \eqref{T00a}. For the first term inside the square brackets we take
\begin{equation}
f(u)=K_{1/2-im\alpha}(i\gamma\eta)K_{1/2-im\alpha}(-i\gamma\eta),
\end{equation}
and for the second one we take
\begin{equation}
f(u)=\gamma^2 K_{1/2-im\alpha}(i\gamma\eta)K_{1/2-im\alpha}(-i\gamma\eta),
\end{equation}
considering $g(u)=1$ for both terms. 
The contribution induced by the cosmic string and the curvature of the dS spacetime, taking the first term on the right-hand side of \eqref{AP}, is given by
\begin{align}
 \langle 0| T_0^0| 0 \rangle_s^{dS}&=-\frac{q\eta^5}{4\pi^{3}\alpha^4}\sum_j\int_0^{\infty}dp\,p
[J^2_{\beta_1}(pr)+J^2_{\beta_2}(pr)]\left\{\hat{S}_\eta\int_0^\infty dk\,  K_{1/2-im\alpha}(i\gamma\eta) K_{1/2-im\alpha}(-i\gamma\eta)\right.\nonumber\\
&\left.  +
4 \int_0^\infty dk \ \gamma^2K_{1/2-im\alpha}(i\gamma\eta)K_{1/2-im\alpha}(-i\gamma\eta)\right\} .
\end{align}
Thanks to the representation \eqref{MacRep} the integration over the variable $k$ can be performed
following a similar procedure adopted in the Appendix A of the Ref. \cite{BezerradeMello:2010ci}. As a result we have
 \begin{align}
 \langle 0| T_0^0| 0 \rangle_s^{dS}&=-\frac{q(\eta/r)^5}{2^{3/2}\pi^{5/2}\alpha^4}\int_0^\infty dx\,x^{3/2}e^{-x}
 \mathcal{J}(q,a_0,x)\, \hat{F}_u \, e^u \, K_{1/2-im\alpha}(u)|_{u=x\eta^2/r^2},
 \end{align}
 where we have defined the operator
 \begin{equation}
 \hat{F}_u=u\, \partial_u^2+(3/2-2u)\, \partial_u -2+\frac{im\alpha(1/2-im\alpha)}{u}.
 \label{Fu}
 \end{equation}
 From the properties of the MacDonald functions, one can see that
 \begin{equation}
 \hat{F}_u e^u K_{1/2-im\alpha}(u)=-\frac{1}{2}e^u\, [K_{1/2-im\alpha}(u)+[K_{-1/2+im\alpha}(u)].
 \label{FK}
 \end{equation}
Taking into account the previous results we find
 \begin{align}
  \langle 0| T_0^0| 0 \rangle_s^{dS}&=\frac{q\alpha^{-4}}{2^{3/2}\pi^{5/2}}\int_0^\infty dz z^{3/2}e^{z(1-r^2/\eta^2)}\, {\rm Re}\,
  [K_{1/2-im\alpha}(z)] \mathcal{J}(q,a_0,zr^2/\eta^2),
 \end{align}
 where $\mathcal{J}(q,a_0,zr^2/\eta^2)$ is given by \eqref{representation}.
Following a similar procedure as in \eqref{nondiv} to extract the dS part for the FC and using \eqref{leg1}
 and \eqref{leg2}, we find the final expression for the VEV of the energy density induced by the cosmic string in dS space
 \begin{align}
   \langle 0| T_0^0| 0 \rangle_s&=\frac{1}{\pi^{2}\alpha^4}\sqrt{\frac{2}{\pi}}\left[\sum_{k=1}^p
   (-1)^k\cos(2k\pi a_0)\cos(k\pi/q) \, \mathcal{M}_0(u_{0,k})\right.\nonumber\\
   &\left.+\frac{q}{\pi}\int_0 ^\infty dx \frac{\sinh (x) h(q,a_0,2x)}{\cosh(2qx)-\cos(q\pi)}
   \mathcal{M}_0(u_{0,x})\right],
 \end{align}
with the notation
\begin{align}
\mathcal{M}_l(u_{l,y})={\rm Re}\left[\mathcal{F}_l(u_{l,y})\right],
\label{MLdefinition}
\end{align}
where we have defined $\mathcal{F}_l$ and $u_{l,y}$ in \eqref{Gfunction}-\eqref{udefinition}. We note that the energy density induced by the cosmic string is an even function of $a_0$ and depends on the ratio $r/\eta$.

Now, we focus on the compactification contribution to the VEV of the energy density.
From the second term in the Abel Plana summation formula \eqref{AP} and considering again
\eqref{Euler}, we get
\begin{align}
\langle 0| T_0^0| 0 \rangle_c&=-\frac{iq\eta^5}{8\pi^3\alpha^4}\sum_j \int_0^\infty dp\,p\, 
[J^2_{\beta_1}(pr)+J^2_{\beta_2}(pr)]\int_p^\infty du K_{1/2-im\alpha}(\eta \lambda)\nonumber\\
&\times\left[ \hat{S}_\eta- 4\lambda^2\right]
\left[ K_{1/2-im\alpha}(e^{i\pi}\eta \lambda)-K_{1/2-im\alpha}(e^{-i\pi}\eta\lambda) \right] \sum_{\chi=\pm 1}\left(e^{Lu+2\pi i\chi \tilde{\beta}}-1\right)^{-1}.
\end{align}
In the above integral we considered $\lambda=\sqrt{u^2-p^2}$, as before.
By using again the Eqs. \eqref{Mac1} and \eqref{Mac2} besides the expansion $(e^{u}-1)^{-1}=\sum_{l=1}^{\infty}e^{-lu}$,
we obtain
\begin{align}
\langle 0| T_0^0| 0 \rangle_c&=-\frac{iq\eta^5}{4\pi^2\alpha^4}\sum_{l=1}^{\infty}\cos(2\pi l\tilde{\beta})
\sum_j \int_0^\infty dp \, p \,[J^2_{\beta_1}(pr)+J^2_{\beta_2}(pr)]\int_0^\infty \frac{d\lambda}{\sqrt{\lambda^2+p^2}}\nonumber\\
&\times\left[\lambda\hat{S}(\eta,\alpha)-4\lambda^3\right]K_{1/2-im\alpha}(\eta\lambda)
\left[I_{1/2-im\alpha}(\eta\lambda)+I_{-1/2+im\alpha}(\eta\lambda)\right] e^{-lL\sqrt{\lambda^2+p^2}}.
\end{align}
One can use the identity \eqref{IntRep} which allows to solve the integration over $p$ as well as the 
integration over $\lambda$ knowing $\lambda^3e^{-\lambda^2s^2}=-\lambda\partial_{s^2}e^{-\lambda^2s^2}$.
Doing so and considering similar transformations
that we have done in the evaluation of the contribution induced by the compactification for the FC, we find
\begin{align}
   \langle 0| T_0^0| 0 \rangle_c&=-\frac{q\eta^5}{4\pi^{5/2}\alpha^4}\sum_{l=1}^\infty \cos(2\pi l\tilde{\beta})
   \int_0^\infty ds s^{-6}e^{-\frac{l^2L^2}{4s^2}-\frac{r^2}{2s^2}}\mathcal{J}(q,a_0,r^2/(2s^2))\nonumber\\
   &\times \hat{F}_u \, e^u \, K_{1/2-im\alpha(u)}|_{u=\eta^2/2s^2},
\end{align}
where the operator $\hat{F}_u$ was defined in the Eq. \eqref{Fu}.
The next step, we consider \eqref{FK} and \eqref{representation} to obtain the final expression for the energy
density induced by the compactification which is given by
\begin{align}
\langle 0| T_0^0| 0 \rangle_c&=\frac{2^{3/2}}{\pi^{5/2}\alpha^4}\sum_{l=1}^\infty \cos(2\pi l\tilde{\beta}) \left[\sideset{}{'}\sum_{k=0}^p(-1)^k\cos(2k\pi a_0)\cos(k\pi/q) \, \mathcal{M}_l(u_{l,k})\right.\nonumber\\
&+\left. \frac{q}{\pi}\int_0 ^\infty dx \frac{\sinh (x) h(q,a_0,2x)}{\cosh(2qx)-\cos(q\pi)} \, \mathcal{M}_l(u_{l,x})\right],
\end{align}
where the prime on the summation means that the $k=0$ term has the coefficient $1/2$.

\subsection{Radial stress}
Now, let us consider the VEV of the radial part of the stress-energy tensor. Taking into account $\Gamma_r$ in the Eq. \eqref{EMT} as well as considering the modes \eqref{modes}, we have
\begin{align}
\langle 0|T_1^1|0\rangle&=\frac{q\eta^5}{\pi^2\alpha^4 L}\sum_{l=-\infty}^{\infty}\int_0^\infty dp\,p^3\sum_j\epsilon_j
[J_{\beta_1}(pr)J'_{\beta_2}(pr)-J'_{\beta_1}(pr)J_{\beta_2}(pr)]\nonumber\\
&\times {\rm Re} \, [K_{1/2-im\alpha}(i\gamma\eta)K_{1/2-im\alpha}(-i\gamma\eta)],
\end{align}
where the prime in the Bessel functions means derivative with respect to the argument.
Using the recurrent relation \eqref{bess-rec} for the Bessel functions we obtain
\begin{align}
\langle 0|T_1^1|0\rangle&=\frac{q\eta^5}{\pi^2\alpha^4 L}\sum_{l=-\infty}^{\infty}\int_0^\infty dp\,p^3\sum_j
[J^2_{\beta_1}(pr)+J^2_{\beta_2}(pr)-\frac{2\beta_1+\epsilon_j}{pr}J_{\beta_1}(pr)J_{\beta_2}(pr)]\nonumber\\
&\times{\rm Re} \, [K_{1/2-im\alpha}(i\gamma\eta)K_{1/2-im\alpha}(-i\gamma\eta)].
\end{align}
We shall use the Abel-Plana summation formula to solve the summation over $l$ which allows us to decompose the radial stress as \eqref{decompositionEMT}.
We can use the \eqref{AP} with $g(u)=1$ and $f(u)$ being the real part of \eqref{ufunction}.
From the first term of the right-hand side of the Abel-Plana formula \eqref{AP}, the contribution coming from
the curvature of the dS spacetime and the cosmic string can be written as
\begin{align}
\langle 0|T_1^1|0\rangle_{s}^{dS}&=\frac{q\eta^5}{\pi^3\alpha^4}\sum_j \int_0^\infty dp\,p^3 \int_0^\infty dk \ {\rm Re}\,
[K_{1/2-im\alpha}(i\gamma\eta)K_{1/2-im\alpha}(-i\gamma\eta)]\nonumber\\
&\times [J^2_{\beta_1}(pr)+J^2_{\beta_2}(pr)-\frac{2\beta_1+\epsilon_j}{pr}J_{\beta_1}(pr)J_{\beta_2}(pr)].
\end{align}
Combining the relations \eqref{MacRep}, \eqref{bess-mod} and \eqref{rep-mod2}
where $x=r^2/(4u\eta^2\sinh^2y)$, the final expression for the radial stress induced by the string and the dS curvature is given by
\begin{eqnarray}
\langle 0|T_1^1|0\rangle_{s}^{dS}&=&\frac{q\alpha^{-4}}{2^{3/2}\pi^{5/2}}\int_0^{\infty}dz\,z^{3/2}
e^{z(1-r^2/\eta^2)}{\rm Re}\, [K_{1/2-im\alpha}(z)]\mathcal{J}(q,a_0,zr^2/\eta^2).
\label{T1}
\end{eqnarray}
From this expression we can note that
\begin{equation}
\langle 0|T_1^1|0\rangle_{s}^{dS}=\langle 0|T_0^0|0\rangle_{s}^{dS}.
\end{equation}

Now, for the radial stress induced by the compactification, considering the second term on the right-hand side of the Abel-Plana summation formula \eqref{AP} we have
\begin{align}
\langle 0|T_1^1|0\rangle_c&=\frac{q\eta^5}{\pi^2\alpha^4}\sum_{l=1}^\infty \cos(2\pi l\tilde{\beta})\sum_j\int_0 ^\infty
dp\,p^3 [J^2_{\beta_1}(pr)+J^2_{\beta_2}(pr)-\frac{2\beta_1+\epsilon_j}{pr}J_{\beta_1}(pr)J_{\beta_2}(pr)]\nonumber\\
&\times \int_0^\infty \frac{d\lambda \lambda}{\sqrt{\lambda^2+p^2}}{\rm Re}\left\{K_{1/2-im\alpha}(\eta\lambda)
[I_{1/2-im\alpha}(\eta\lambda)+I_{-1/2+im\alpha}(\eta\lambda)]\right\}e^{-lL\sqrt{\lambda^2+p^2}},\nonumber\\
\end{align}
where we have used the expansion $\left(e^u-1\right)^{-1}=\sum_{l=1}^{\infty}e^{-lu}$ besides the Eqs. \eqref{Mac1} and \eqref{Mac2}.
Considering the integral
representation \eqref{IntRep}, the integral over $\lambda$ can be evaluated directly and the result is
\begin{align}
\langle 0|T_1^1|0\rangle_c&=\frac{q\eta^5}{\pi^{5/2}\alpha^4}\sum_{l=1}^\infty \cos(2\pi l\tilde{\beta})\sum_j
\int_0^\infty ds\,  \frac{e^{-l^2L^2/(4s^2)+\eta^2/(2s^2)}}{s^2}\, {\rm Re}\, [K_{1/2-im\alpha}(\eta^2/2s^2)]\nonumber\\
&\times\int_0^\infty dp\,p^3 e^{-p^2s^2}[J^2_{\beta_1}(pr)+J^2_{\beta_2}(pr)-\frac{2\beta_1+\epsilon_j}{pr}J_{\beta_1}(pr)J_{\beta_2}(pr)].
\end{align}
To solve the integral over $p$ we make similar considerations done for the contribution induced by the curvature of
the dS spacetime and the cosmic string which results in
\begin{align}
\langle 0|T_1^1|0\rangle_c&=\frac{q\alpha^{-4}}{2^{1/2}\pi^{5/2}}\sum_{l=1}^\infty \cos(2\pi l\tilde{\beta})
\int_0^\infty dz\,z^{3/2}e^{z(1-r^2/\eta^2-l^2L^2/(2\eta^2))}\, {\rm Re}\, [K_{1/2-im\alpha}(z)]\nonumber\\
&\times \mathcal{J}(q,a_0,zr^2/\eta^2).
\end{align}
As in the case of the string contribution, we note that 
\begin{equation}
\langle 0|T_1^1|0\rangle_c=\langle 0|T_0^0|0\rangle_c.
\end{equation}

\subsection{Azimuthal stress}
\label{T22sec}
Our next step is the evaluation of the azimuthal stress. In order to proceed we take into account
that \cite{BezerradeMello:2010ci}
\begin{equation}
\Gamma_2=-\frac{1}{2\alpha}\gamma^0\gamma_2+\frac{1-q}{2}\gamma^{(1)}\gamma^{(2)}.
\end{equation}
Using the negative energy eingenfunction in the
mode sum \eqref{EMT}, one finds
\begin{align}
\langle 0|T_2^2|0\rangle&=\frac{2q^2\eta^5}{\pi^2\alpha^4L r^2}\sum_{l=-\infty}^{\infty}\sum_j(j+a)
\left(\epsilon_j \beta_1-\frac{r}{2}\partial_r\right)\int_0^\infty dp \, p\,  J^2_{\beta_1}(pr)\nonumber\\
&\times{\rm Re}\, [K_{1/2-im\alpha}(i\gamma\eta)K_{1/2-im\alpha}(-i\gamma\eta)],
\end{align}
where we have used the relation \eqref{bess-2}.

Again, using the Abel-Plana summation formula, considering $g(u)=1$ and $f(u)$ as the
real part of \eqref{ufunction}, we are able to decompose this component of the energy-momentum
tensor into two contributions. For the contribution induced by the curvature of the dS spacetime and the cosmic string, and
considering similar transformations previously done, we have
\begin{align}
\langle 0|T_2^2|0\rangle_{s}^{dS}&=\frac{q^2}{2^{1/2}\pi^{5/2}\alpha^4}\sum_j(j+a)\int_0^\infty
dz\,z^{3/2}e^z \, {\rm Re}\, [K_{1/2-im\alpha}(z)]\nonumber\\
&\times\left(\frac{\epsilon_j\beta_1}{x}-\partial_x\right)e^{-x}I_{\beta_1}(x)|_{x=zr^2/\eta^2}.
\end{align}
Following the representations \eqref{rep-mod} and \eqref{rep-mod2}, one obtains
\begin{align}
\langle 0|T_2^2|0\rangle_{s}^{dS}&=\frac{q}{2^{1/2}\pi^{5/2}\alpha^4}\int_0^\infty dz\,z^{3/2}
e^{z(1-r^2/\eta^2)}{\rm Re}[K_{1/2-im\alpha}(z)]\nonumber\\
&\times \left(z\partial_z-zr^2/\eta^2+1/2\right)\sum_j[I_{\beta_1}(zr^2/\eta^2)+I_{\beta_2}(zr^2/\eta^2)].
\end{align}
knowing $q(j+a)=\epsilon_j\beta_1+1/2$. Writing the parameter $a$ as indicated by \eqref{a0} and comparing the above equation
with the Eq. \eqref{T1}, we note that following relation holds
\begin{equation}
\langle 0|T_2^2|0\rangle_{s}^{dS}=\partial_r\left[r \, \langle 0|T_1^1|0\rangle_{s}^{dS}\right].
\end{equation}
Using the corresponding radial stress, and extracting the divergent part, we find the azimuthal one in the following form
\begin{align}
   \langle 0| T_2^2| 0 \rangle_s&=\frac{1}{2^{1/2}\pi^{5/2}\alpha^4} \left[\sum_{k=1}^p
   (-1)^k\cos(2k\pi a_0)\cos(k\pi/q) \, \mathcal{M}^{(\phi)}_0\left(u_{0,k},\frac{r}{\eta}s_k\right)\right.\nonumber\\
   &\left. +\frac{q}{\pi}\int_0 ^\infty dx \frac{\sinh (x) h(q,a_0,2x)}{\cosh(2qx)-\cos(q\pi)}\,
   \mathcal{M}^{(\phi)}_0\left(u_{0,x},\frac{r}{\eta}c_x\right)\right],
   \nonumber\\
   \label{T22S}
 \end{align}
where we define the notation
\begin{align}
\mathcal{M}_l^{(\phi)}(u_{l,y},v)&=\frac{\sqrt{\pi}}{2^{5/2}} \, {\rm Re} \left\{\frac{\Gamma(3-im\alpha)\Gamma(2+im\alpha)}{u_{l,y}^{6-im\alpha}}
\left[u_{l,y}^2\left(\frac{l^2L^2}{2\eta^2}+2(2im\alpha-5)v^2-1\right)\right.\right.\nonumber\\
&\left.\left.\times F\left(a,b;c;d_{l,y}\right)+\frac{2}{3}(3-im\alpha)(4-im\alpha)
v^2\right.\left.  F\left(a+1,b+1;c+1;d_{l,y}\right)\right]\right\}.
\end{align}
taking $l=0$.
 
Considering the second term in the Abel-Plana summation formula
and doing similar transformations, for the contribution of the azimuthal stress induced by the compactification we find
\begin{align}
\langle 0| T_2^2| 0 \rangle_c&=\frac{2^{1/2}q}{\pi^{5/2}\alpha^4}\sum_{l=1}\cos(2\pi \tilde{\beta})\int_0^\infty
dz\,z^{3/2}e^{-z(l^2L^2/(2\eta^2)+r^2/\eta^2)}{\rm Re}[K_{1/2-im\alpha}(z)]\nonumber\\
&\times \left(z\partial_z-zr^2/\eta^2+1/2\right)\sum_j[I_{\beta_1}(zr^2/\eta^2)+I_{\beta_2}(zr^2/\eta^2)].
\end{align}
Again, it is easy to show that
\begin{equation}
\langle 0|T_2^2|0\rangle_{c}=\partial_r\left[ r \, \langle 0|T_1^1|0\rangle_{c}\right],
\end{equation}
which leads to the final expression for the azimuthal stress induced by the compactification as
\begin{align}
\langle 0| T_2^2| 0 \rangle_c&=\frac{2^{1/2}}{\pi^{5/2}\alpha^4}\sum_{l=1}^\infty \cos(2\pi l\tilde{\beta}) \left[\sideset{}{'}
\sum_{k=0}^p(-1)^k\cos(2k\pi a_0)\cos(k\pi/q) \, \mathcal{M}^{(\phi)}_l\left(u_{l,k},\frac{r}{\eta}s_k\right)\right.\nonumber\\
&\left. +\frac{q}{\pi}\int_0 ^\infty dx\, \frac{\sinh (x) h(q,a_0,2x)}{\cosh(2qx)-\cos(q\pi)}
\mathcal{M}^{(r)}_l\left(u_{l,x},\frac{r}{\eta}c_x\right)\right].
\label{T22C}
\end{align}
Taking into account \eqref{T22S} and \eqref{T22C}, it is possible to write an expression for the total azimuthal stress. This expression reads
 \begin{align}
\langle 0| T_2^2| 0 \rangle&=\frac{2^{1/2}}{\pi^{5/2}\alpha^4}\sum_{l=0}^\infty \cos(2\pi l\tilde{\beta}) \left[\sideset{}{'}\sum_{k=0}^p(-1)^k\cos(2k\pi a_0)\cos(k\pi/q) \, \mathcal{M}^{(\phi)}_l\left(u_{l,k},\frac{r}{\eta}s_k\right)\right.\nonumber\\
&\left. +\frac{q}{\pi}\int_0 ^\infty dx\, \frac{\sinh (x) h(q,a_0,2x)}{\cosh(2qx)-\cos(q\pi)}
\mathcal{M}^{(\phi)}_l\left(u_{l,x},\frac{r}{\eta}c_x\right)\right],
\label{T22total}
\end{align}
where the contribution induced by the cosmic string is the $l=0$ term of the above equation with the coefficient $1/2$.

\subsection{Axial stress}
\label{T33sec}
For the axial stress, considering the equation \eqref{EMT} after inserting the expression for the eigenfunctions, one finds
 \begin{align}
 \langle 0| T_3^3| 0 \rangle&=\frac{q\eta^5}{\alpha^4\pi^2 L}\sum_{l=-\infty}^{\infty} \sum_j\int_0^\infty dp\,p  
 [J^2_{\beta_1}(pr)+J^2_{\beta_2}(pr)] \,k_l^2\nonumber\\
&\times {\mathrm{Re}}\, [K_{1/2-im\alpha}(i\gamma\eta)K_{1/2-im\alpha}(-i\gamma\eta)].
 \end{align}
 To solve the summation over $l$ we consider the Abel-Plana summation formula \eqref{AP} with the functions $g(u)=k^2$ and
 $f(u)$ given by the real part of \eqref{ufunction}.
 For the contribution induced by the cosmic string and the de Sitter spacetime, using the first term of
 \eqref{AP} and the representation \eqref{MacRep}, after similar transformations done previously one finds
 \begin{align}
  \langle 0| T_3^3| 0 \rangle_s^{dS}&=\frac{q\alpha^{-4}}{2^{3/2}\pi^{5/2}}\int_0^\infty dz\,z^{3/2}e^{z(1-r^2/\eta^2)}{\rm Re}
  [K_{1/2-im\alpha}(z)] \mathcal{J}(q,a_0,zr^2/\eta^2),
 \end{align}
 where we can note that 
 \begin{equation}
 \langle 0| T_3^3| 0 \rangle_s^{dS}=\langle 0| T_0^0| 0 \rangle_s^{dS}.
 \end{equation}
 
The axial stress induced by the compactification is obtained from the second term in \eqref{AP}. By using this formula
and taking into account similar transformations done previously for the contributions induced by the compactification we obtain
\begin{align}
\langle 0| T_3^3| 0 \rangle_c&=-\frac{q\eta^5}{\alpha^4\pi^2}\sum_{l=1}^\infty \cos(2\pi l\tilde{\beta})\sum_j\int_0^\infty
dp\,p\, [J^2_{\beta_1}(pr)+J^2_{\beta_2}(pr)]\int^\infty_p du\,u^2 e^{-lLu}\nonumber\\
&\times {\rm Re}\left\{K_{1/2-im\alpha}(\eta \lambda)[I_{1/2-im\alpha}(\eta \lambda)+
I_{-1/2+im\alpha}(\eta \lambda)]\right\},
\end{align}
where again we have considered \eqref{Euler}.
Knowing $u^2e^{-lLu}=l^{-2}\partial^2_L e^{-lLu}$ and using the representation \eqref{IntRep} we are able
to solve the integrals over $p$ and $u$ which gives
\begin{align}
\langle 0| T_3^3| 0 \rangle_c&=\frac{q\eta^5}{4\alpha^4\pi^{5/2}}\sum_{l=1}^\infty \cos(2\pi l\tilde{\beta})
\int_0^\infty ds\left(1-\frac{l^2L^2}{2s^2}\right)\frac{e^{-(l^2L^2/2  + r^2 -\eta^2)/(2s^2)}}{s^6}
\nonumber\\
&\times{\rm Re}[K_{1/2-im\alpha}(\eta^2/2s^2)] \, \mathcal{J}(q,a_0,r^2/2s^2).
\end{align}
By making the change of variables $z=\eta^2/2s^2$ and also using the representation \eqref{representation} we
note that the following relation holds
\begin{equation}
\langle 0| T_3^3| 0 \rangle_c=\partial_L\left[L \, \langle 0| T_1^1| 0 \rangle_c\right].
\end{equation}
Taking into the consideration the expression for the radial stress induced by the compactification,
the axial stress is given by
\begin{align}
\langle 0| T_3^3| 0 \rangle_c&=\frac{2^{3/2}}{\pi^{5/2}\alpha^4}\sum^\infty_{l=1}\cos(2\pi l\tilde{\beta})
 \left[\sideset{}{'}\sum_{k=0}^p(-1)^k\cos(2k\pi a_0)\cos(k\pi/q) \, \mathcal{M}^{(z)}_l\left(u_{l,k},\frac{r}{\eta}s_k\right)\right.\nonumber\\
&\left. +\frac{q}{\pi}\int_0^\infty
dx\, \frac{\sinh(x)h(q,a_0,2x)}{\cosh(2qx)-\cos(q\pi)}\mathcal{M}^{(z)}_l\left(u_{l,x},\frac{r}{\eta}c_x\right)\right],
\label{T33C}
\end{align}
with the notation
\begin{align}
\mathcal{M}_l^{(z)}(u_{l,y},v)&=\frac{\sqrt{\pi}}{2^{5/2}} \, {\rm Re} \left\{\frac{\Gamma(3-im\alpha)\Gamma(2+im\alpha)}{u_{l,y}^{6-im\alpha}}
\left[u_{l,y}^2\left((2im\alpha-5)\frac{l^2L^2}{2\eta^2}+2v^2-1\right)\right.\right.\nonumber\\
&\left.\left.\times F\left(a,b;c;d_{l,y}\right)+\frac{1}{3}(3-im\alpha)(4-im\alpha)
\frac{l^2L^2}{2\eta^2}\right.\left.  F\left(a+1,b+1;c+1;d_{l,y}\right)\right]\right\}.
\end{align}

\section{Properties of the VEV of the energy-momentum tensor}
\label{EMTsec2}
In this section, we plan to analyze some properties of the VEVs of the energy-momentum tensor. As we have shown, the VEVs of the energy-momentum tensor can be decomposed into a contribution induced by the curvature of the de Sitter spacetime and the cosmic string, and another one induced by the compactification. The first contribution,
after removing the pure dS part, can be written in a compact form as (no summation over $\mu$)
\begin{align}
   \langle 0| T_\mu^\mu| 0 \rangle_s&=\frac{2^{1/2}}{\pi^{5/2}\alpha^4}\left[\sum_{k=1}^p
   (-1)^k\cos(2k\pi a_0)\cos(k\pi/q) \, \mathcal{M}_0(u_{0,k})\right.\nonumber\\
   &\left.+\frac{q}{\pi}\int_0 ^\infty dx \frac{\sinh (x) h(q,a_0,2x)}{\cosh(2qx)-\cos(q\pi)} \,
   \mathcal{M}_0(u_{0,x})\right],
   \label{TS1}
 \end{align}
for $\mu=t, r, z$ where the functions $\mathcal{M}_0(u_{0,y})$ were defined in \eqref{MLdefinition}, and also \eqref{Gfunction}-\eqref{udefinition}.
This contribution of the energy-momentum tensor is an even function of $a_0$.
Considering a massless fermionic field, the above expression is simplified to
\begin{align}
   \langle 0| T_\mu^\mu| 0 \rangle_s&=\frac{1}{4\pi^2\alpha^4}\left[\sum_{k=1}^p
   (-1)^k\, \frac{\cos(2k\pi a_0)\cos(k\pi/q)}{u_{0,k}^3} \, F\left(2,3/2;3;1-u^{-2}_{0,k}\right)
  \right.\nonumber\\
   &\left.+\frac{q}{\pi}\int_0 ^\infty dx\, \frac{\sinh (x) h(q,a_0,2x)}{\cosh(2qx)-\cos(q\pi)} \,
   u_{0,x}^{-3} \, F\left(2,3/2;3;1-u^{-2}_{0,x}\right)\right].
   \label{TS1massless}
 \end{align}

Now, let us consider some asymptotic behavior of the expression \eqref{TS1}. Near the string, $r/\eta\ll 1$, we
proceed in a similar way as was done in the string part of the FC. First, we write \eqref{TS1} as
\begin{align}
\langle 0| T_\mu^\mu| 0 \rangle_s&=\frac{2^{1/2}}{\pi^{5/2}\alpha^4}\left[\sum_{k=1}^p
   (-1)^k \cos(2k\pi a_0)\cos(k\pi/q)\int_0^\infty dz\, z^{3/2}e^{-z(2r^2s_k^2/\eta^2 -1)}\right.\nonumber\\
   &\left.+\frac{q}{\pi}\int_0 ^\infty dx\,  \frac{\sinh (x) h(q,a_0,2x)}{\cosh(2qx)-\cos(q\pi)}
   \int_0^\infty dz\, z^{3/2}e^{-z(2r^2c_x^2/\eta^2 -1)}\right]\, {\rm Re}[K_{1/2-im\alpha}(z)].
\end{align}
For the region near the string, the main contribution to the integral over $z$ comes from the regions
near the upper limit of the integration, and we can use the expression for the Macdonald function considering
large arguments. To the leading order we obtain
\begin{align}
\langle 0| T_\mu^\mu| 0 \rangle_s&\approx\frac{1}{2\pi^2}\left(\frac{\eta}{\alpha r}\right)^4
\left[\sum_{k=1}^p\, (-1)^k \, \frac{\cos(2k\pi a_0)\cos(k\pi/q)}{s_k^{4}}
+\frac{q}{\pi}\int_0 ^\infty dx \frac{c_x^{-4}\sinh (x) h(q,a_0,2x)}{\cosh(2qx)-\cos(q\pi)}\right],
\label{TS1R0}
\end{align}
where we note a divergence with a fourth power of the proper distance.

On the other hand, for regions where $r/\eta\gg 1$, we have 
\begin{align}
\mathcal{M}_0(u_{0,y})&\approx {\rm Re}\left[\frac{\Gamma(1/2-im\alpha)\Gamma(2+im\alpha)}{2^{2im\alpha+5/2}
(u_{0,y}+1)^{2+im\alpha}}\right],
\end{align}
and to the leading order, Eq. \eqref{TS1} reads
\begin{align}
\langle 0| T_\mu^\mu| 0 \rangle_s&\approx \frac{2^{1/2}}{\pi^{5/2}}\left(\frac{\eta}{\alpha r}\right)^{4} \,{\rm Re}
\left\{\frac{\Gamma(1/2-im\alpha)\Gamma(2+im\alpha)}{2^{2im\alpha+5/2}}\left(\frac{\eta}{r}\right)^{2im\alpha}
\left[\sum_{k=1}^p(-1)^k\frac{\cos(2k\pi a_0)\cos(k\pi/q)}{s_k^{4+2im\alpha}}\right.\right.\nonumber\\
&\left.\left.+\frac{q}{\pi}\int_0 ^\infty dx \frac{\sinh (x) h(q,a_0,2x)}{\cosh(2qx)-\cos(q\pi)}c_x^{-4-2im\alpha}\right]\right\}.
\label{TS1largeR}
\end{align}

\begin{figure}[h]
	\centering
	{\includegraphics[width=0.48\textwidth]{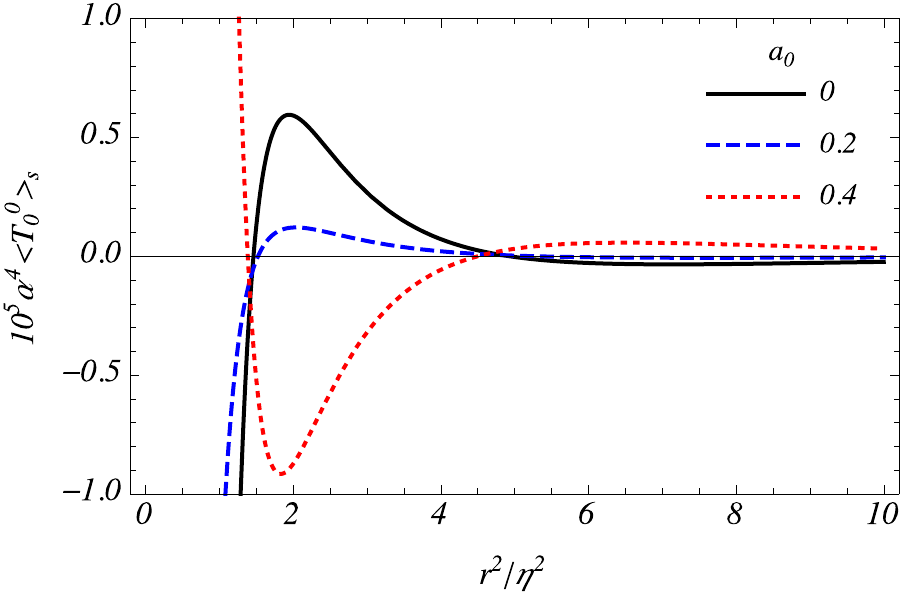}}
	\hfill
	{\includegraphics[width=0.486\textwidth]{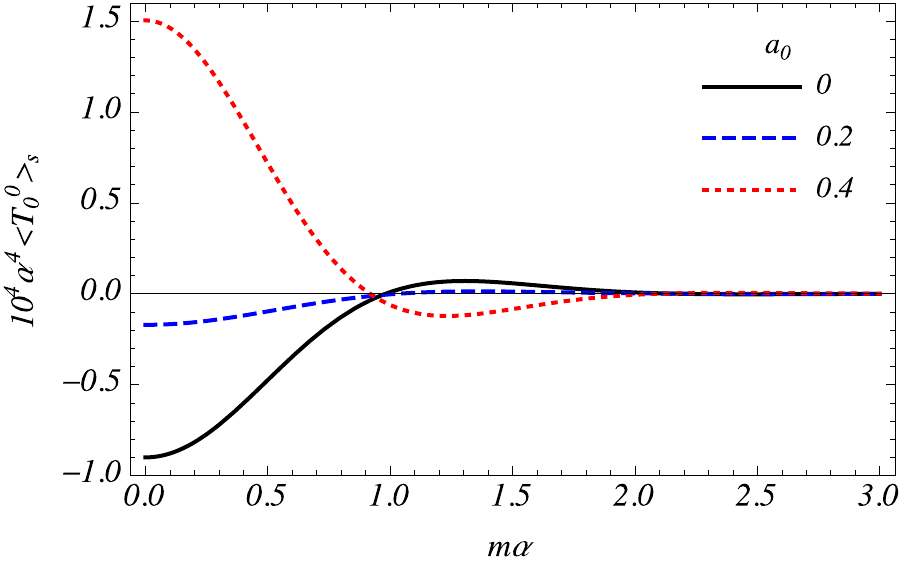}}
	\caption{Energy density and azimuthal stress induceds by the magnetic flux and planar angle deficit as a function of $r^2/\eta^2$ (left plots) for $m\alpha=2$ and the mass of the field (right plots) considering $r/\eta=2$. In both plots $q=1.5$.}
	\label{fig03}
\end{figure}

In the Fig. \ref{fig03} we show the behavior of the string part of the energy density as function of $r^2/\eta^2$ (left plot) and of $m\alpha$ (right plot). As in the case of the string part of FC, the behavior of the string part of the energy density is highly affected by the value of the azimuthal magnetic flux. The turning point is around $a_0\approx 0.23$ for the chosen parameters in the figure.

In a similar way, the contribution due to the compactification, can be written as (no summation over $\mu$)
\begin{align}
\langle 0| T_\mu^\mu| 0 \rangle_c& =\frac{2^{3/2}}{\pi^{5/2}\alpha^4}\sum_{l=1}^{\infty}\cos(2\pi l\tilde{\beta})
\left[\sideset{}{'}\sum_{k=0}^p(-1)^k\cos(2k\pi a_0)\cos(k\pi/q)
\, \mathcal{M}_l(u_{l,k})\right.\nonumber\\
&+\left. \frac{q}{\pi}\int_0 ^\infty dx \frac{\sinh (x) h(q,a_0,2x)}{\cosh(2qx)-\cos(q\pi)} \, \mathcal{M}_l(u_{l,x})\right],
\label{TC1}
\end{align}
for $\mu=t,r$. The previous expression is an even function of the parameters $\tilde{\beta}$ and $a_0$.
Considering $r/\eta \ll1$, the leading term of the above expression reads
\begin{align}
\langle 0| T_\mu^\mu| 0 \rangle_c&\approx\frac{q \cos[q\pi(1/2+|a_0|)]\Gamma(1/2-q(1/2-|a_0|))}{\pi^{7/2}\alpha^4}
\left(\frac{r}{\sqrt{2}\eta}\right)^{q(1-2|a_0|)-1}\nonumber\\
&\times\sum_{l=1}^{\infty}\cos(2\pi l\tilde{\beta}) \, {\rm Re}
\left\{\frac{\Gamma(2+im\alpha)\Gamma(3/2-im\alpha+q(1-2|a_0|))}{2^{2+im\alpha}\Gamma(2-im\alpha)
(lL/(\sqrt{2}\eta)^{3-im\alpha+q(1-2|a_0|)})}\right\}.
\label{TC1R0}
\end{align}
This expression is divergent if $|a_0|>(1-1/q)/2$ and is finite if $|a_0|\leq(1-1/q)/2$ in the limit $r=0$.
For $a_0=0$ the above expression vanishes on the string's core.

The Eq. \eqref{TC1} can be decomposed as
\begin{equation}
\langle 0| T_\mu^\mu| 0 \rangle_c=\langle 0| T_\mu^\mu| 0 \rangle_c^{(0)}+\langle 0| T_\mu^\mu| 0 \rangle_c^{(q,a_0)}.
\end{equation}
The first term is the $k=0$ contribution with the coefficient $1/2$ in \eqref{TC1}. This is a pure topological term,
independent of the radial coordinate, the azimuthal magnetic flux and the conical defect. This contribution is induced only by
the compactification which is given by
\begin{equation}
\langle 0| T_\mu^\mu| 0 \rangle_c^{(0)}=\frac{2^{1/2}}{\pi^{5/2}\alpha^4}\sum_{l=1}^{\infty}\cos(2\pi l\tilde{\beta})
\label{TC1a}
\mathcal{M}_l(u_{l,0}).
\end{equation}
The second contribution on the right-hand side of \eqref{TC1} which depends on the planar angle deficit, magnetic flux and compactification, is given by
\begin{align}
\langle 0| T_\mu^\mu| 0 \rangle_c^{(q,a_0)}& =\frac{2^{3/2}}{\pi^{5/2}\alpha^4}\sum_{l=1}^{\infty}\cos(2\pi l\tilde{\beta})
\left[\sum_{k=1}^p(-1)^k\cos(2k\pi a_0)\cos(k\pi/q)
\, \mathcal{M}_l(u_{l,k})\right.\nonumber\\
&+\left. \frac{q}{\pi}\int_0 ^\infty dx \, \frac{\sinh (x) h(q,a_0,2x)}{\cosh(2qx)-\cos(q\pi)} \, \mathcal{M}_l(u_{l,x})\right].
\label{TC2}
\end{align}

For $L/\eta \gg 1$ and fixed $r/\eta$, to the leading order, we have
\begin{align}
\langle 0| T_\mu^\mu| 0 \rangle_c^{(q,a_0)}&\approx\frac{2^{3/2}}{\pi^{5/2}}\left(\frac{\eta}{\alpha L}\right)^{4}
\sum_{l=1}^{\infty}\cos(2\pi l\tilde{\beta}) \, {\rm Re}\left\{\frac{\Gamma(1/2-im\alpha)\Gamma(2+im\alpha)}
{2^{im\alpha +1/2}}\left(\frac{\eta}{L}\right)^{2im\alpha}\right.\nonumber\\
&\left.\times\left[\sum_{k=1}^p\frac{(-1)^k\cos(2k\pi a_0)\cos(k\pi/q)}
{\left(\frac{l^2}{2}+\frac{2r^2s_k^2}{L^2}\right)^{2+im\alpha}} +
\frac{q}{\pi}\int_0 ^\infty dx \frac{\sinh (x) h(q,a_0,2x)}{\cosh(2qx)-\cos(q\pi)}\right.\right.\nonumber\\
&\left.\left.\times \left(\frac{l^2}{2}+\frac{2r^2c_x^2}{L^2}\right)^{-2-im\alpha}\right]\right\}.
\label{TC2largeL}
\end{align}

\begin{figure}[h]
	\centering
	{\includegraphics[width=0.48\textwidth]{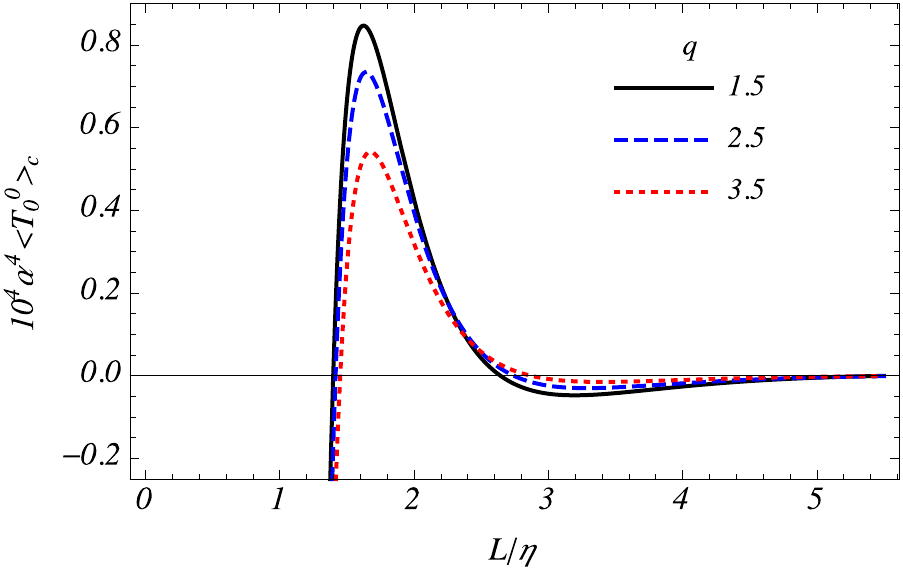}}
	\hfill
	{\includegraphics[width=0.486\textwidth]{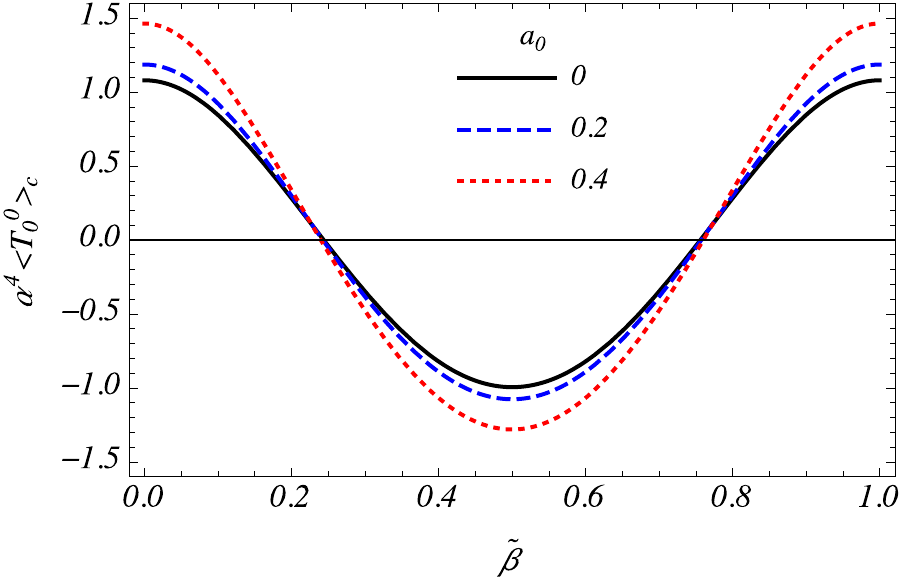}}
	\caption{Energy density induced by the compactification as a function of $L/\eta$ (left plot) for $\tilde{\beta}=r/\eta=a_0=0.25$, considering different
	values of $q$
	and as a function of $\tilde{\beta}$ (right plot) for $L/\eta=0.75$, $r/\eta=0.25$ and $q=1.5$, for three different values of $a_0$.}
	\label{fig04}
\end{figure}

In Fig. \ref{fig04} we show the behavior of the energy density induced by the compactification as a function of $L/\eta$ (left plot) and $\tilde{\beta}$
(right plot). We note that, this contribution of the energy density goes to zero for large values of $L$, as expected.

Now, let us evaluate some special cases of the azimuthal stress given by the Eq. \eqref{T22total}.
For a massless fermion field the azimuthal stress reads
\begin{align}
\langle 0| T_2^2| 0 \rangle&=\frac{1}{2\pi^2\alpha^4}\sum_{l=0}^{\infty}\cos(2\pi l \tilde{\beta})\left[\sideset{}{'}\sum_{k=0}^{p}(-1)^k
\cos(2k\pi\alpha_0)\cos(k\pi/q)\, \mathcal{B}_l\left(u_{l,k},\frac{r}{\eta}s_k\right)\right.\nonumber\\
&\left. +\frac{q}{\pi}\int_0^\infty dx\frac{h(q,a_0,2x)\sinh x}{\cosh(2qx)-\cos(q\pi)}\mathcal{B}_l\left(u_{l,x},\frac{r}{\eta}c_x\right)\right],
\label{T22massless}
\end{align}
with the notation
\begin{align}
\mathcal{B}^{(\phi)}_{l}(u_{l,y},v)&=
\frac{\left(\frac{l^2L^2}{2\eta^2}-10v^2-1\right)}{u_{l,y}^{4}}\, F\left(2,3/2;3;1-u_{l,y}^{-2}\right)+\, \frac{4v^2 }{u_{l,y}^{6}}\, F\left(3,5/2;4;1-u_{l,y}^{-2}\right).
\end{align}
In the absence of the magnetic flux in the azimuthal direction, $a_0=0$, we have
\begin{align}
\langle 0| T_2^2| 0 \rangle&=\frac{2^{1/2}}{\pi^{5/2}\alpha^4}\sum_{l=0}^\infty \cos(2\pi l\tilde{\beta})
\left[\sideset{}{'}\sum_{k=0}^p(-1)^k\cos(k\pi/q) \, \mathcal{M}^{(\phi)}_l\left(u_{l,k},\frac{r}{\eta}s_k\right)\right.\nonumber\\
&\left. +\frac{2q\cos(q\pi/2)}{\pi}\int_0 ^\infty dx \frac{\sinh (x) \sinh(qx)}{\cosh(2qx)-\cos(q\pi)}
\mathcal{M}^{(\phi)}_l\left(u_{l,x},\frac{r}{\eta}c_x\right)\right].
\label{T22a0}
\end{align}

The axial stress induced by the compactification is given by the Eq. \eqref{T33C}. Considering a massless fermion field,
this expression reads
\begin{align}
\langle 0| T_3^3| 0 \rangle_c&=\frac{1}{\pi^{2}\alpha^4}\sum^\infty_{l=1}\cos(2\pi l\tilde{\beta})
 \left[\sideset{}{'}\sum_{k=0}^p(-1)^k\cos(2k\pi a_0)\cos(k\pi/q) \, \mathcal{B}^{(z)}_l\left(u_{l,k},\frac{r}{\eta}s_k\right)\right.\nonumber\\
&\left. +\frac{q}{\pi}\int_0^\infty
dx \frac{\sinh(x)h(q,a_0,2x)}{\cosh(2qx)-\cos(q\pi)}\mathcal{B}^{(z)}_l\left(u_{l,x},\frac{r}{\eta}c_x\right)\right],
\label{T33Cmassless}
\end{align}
with the notation
\begin{align}
\mathcal{B}^{(z)}_l(u_{l,y},v)&=\frac{\left(2v^2-\frac{5l^2L^2}{2\eta^2}-1\right)}{u_{l,y}^{4}}F\left(2,3/2;3,1-u^{-2}_{l,y}\right)
-\frac{2l^2L^2/\eta^2}{u^6_{l,y}}F\left(3,5/2;4,1-u^{-2}_{l,y}\right).
\end{align}
Again, it is easy to see that the axial stress induced by the compactification can be decomposed as
\begin{equation}
\langle 0|T_3^3|0\rangle_c=\langle 0|T_3^3|0\rangle_c^{(0)}+\langle 0|T_3^3|0\rangle_c^{(q,a_0)}.
\label{T33decomposition}
\end{equation}
The first contribution of the above decomposition is the axial stress induced only by the compactification,
being a pure topological term independent of the radial coordinate and the magnetic flux in the
azimuthal direction. This part is the $k=0$ term in \eqref{T33C} which is given by
\begin{align}
\langle 0|T_3^3|0\rangle_c^{(0)}=\frac{2^{1/2}}{\pi^{5/2}\alpha^4}\sum_{l=1}^{\infty}\cos(2\pi l\tilde{\beta})
\mathcal{M}^{(z)}_l\left(u_{l,0},\frac{r}{\eta}s_0\right).
\label{T33C0}
\end{align}
The second term on the right-hand side of \eqref{T33decomposition}, the compactification contribution to the axial stress dependent on
the planar angle deficit and the magnetic flux, has the form
\begin{align}
\langle 0| T_3^3| 0 \rangle_c&=\frac{2^{3/2}}{\pi^{5/2}\alpha^4}\sum^\infty_{l=1}\cos(2\pi l\tilde{\beta})
 \left[\sum_{k=1}^p(-1)^k\cos(2k\pi a_0)\cos(k\pi/q) \, \mathcal{M}^{(z)}_l\left(u_{l,k},\frac{r}{\eta}s_k\right)\right.\nonumber\\
&\left. +\frac{q}{\pi}\int_0^\infty
dx \frac{\sinh(x)h(q,a_0,2x)}{\cosh(2qx)-\cos(q\pi)}\mathcal{M}^{(z)}_l\left(u_{l,x},\frac{r}{\eta}c_x\right)\right],
\label{T33Cqa0}
\end{align}
where we note that the axial stress induced by the compactification is an even function of the parameters
$\tilde{\beta}$ and $a_0$.

Two important relations obeyed by the fermionic energy-momentum tensor are the conservation condition and the trace relation. The latter can be obtained by using the general expression for the energy-momentum tensor, Eq. \eqref{EMT}, combined with the equation of motion. We can start with the trace relation. Because the contribution to the energy-momentum tensor associated with the cosmic string and compactification present no anomalies, the trace relations read
\begin{equation}
\langle 0| T_\mu^\mu| 0 \rangle_s=m\langle \bar{\psi}\psi\rangle_s \ \ , \ \ \langle 0| T_\mu^\mu| 0 \rangle_c=m\langle \bar{\psi}\psi\rangle_c,
\label{tracerelation}
\end{equation}
where we observe that for a massless field the VEV of the energy-momentum tensor is traceless. {\bf As to the conservation condition}, $\nabla_\alpha \langle 0| T_\beta^\alpha| 0 \rangle=0$, in the configuration
that we are studying here, this equation is reduced only to one differential equation
\begin{equation}
\partial_r\left[r \, \langle 0| T_1^1| 0 \rangle\right]=\langle 0| T_2^2| 0 \rangle.
\label{conservation}
\end{equation}
Note that we have already proved this equation during the evaluation of the energy-momentum tensor in the previous sections.

\section{Conclusion}
\label{conc}
In the present paper, we have studied the influence of the combined effects of the spacetime background and the topology
on the FC and the VEV of the energy-momentum tensor associated with a massive spinor field. As the background
geometry we have considered the dS spacetime, maximally symmetric curved space, conformally flat with constant positive curvature, which plays an important role in the quantum field theory and most importantly in cosmology. We have investigated the effect of the presence of the cosmic string in the aforementioned spacetime and a magnetic flux as well as the compactification
of the spatial dimension along the string inducing topological effects in a fermionic system, considering that the fermion field obeys a quasiperiodic condition \eqref{quasiperiodic}.

The FC has been evaluated by using the direct summation over the fermionic modes, prepared in the Bunch-Davies vacuum
state, given by \eqref{modes}. The momentum along the string direction 
becomes discrete due the compactification along this direction and in order to evaluate the summation over this
quantum number, we have employed the Abel-Plana summation formula \eqref{AP}. Consequently, both the FC
and the VEV of the energy-momentum tensor are decomposed into a contribution induced by a cosmic
string in the dS space with no compactification, and another one induced by the compactification.

The string part of the FC is given by \eqref{FCS3}, where this contribution is an even function of the
magnetic flux along the azimuthal direction and depends on the ratio between the radial
and conformal time coordinates, $r/\eta$, which is the proper distance
from the string measured in units of the dS curvature scale $\alpha$. Some particular cases of the FC contribution induced by
the cosmic string in dS spacetime have been considered. 
For points near the string the FC is given by \eqref{FCR0} where the
leading term presents a divergence with the second power of the proper distance and for large distances from the string
the leading term of the FC is given by \eqref{FClargeR}. In Fig. \ref{fig01} we show the influence of the magnetic flux in
the azimuthal direction on the FC as function of $r^2/\eta^2$ and $m\alpha$. The FC induced by the compactification
is given by \eqref{FCC} being an even function of the magnetic flux in the azimuthal and axial directions. This induced contribution of
the FC can be decomposed into two parts as shown in \eqref{FCCdecomp}. The first part, given by \eqref{FCCdecomp1} is
a pure topological term induced only by the compactification and independent of the parameters $q$, $a_0$ and
the radial coordinate. The second part, Eq. \eqref{FCC2}, is the contribution of the FC induced by the compactification, the planar
angle deficit and the magnetic fluxes. For regions near the string, the leading term of the FC induced by the compactification
is given by \eqref{FCCR0}, which is finite on the string if $|a_0|=(1-1/q)/2$, vanishes
for $|a_0|<(1-1/q)/2$ and diverges if $|a_0|>(1-1/q)/2$. We have also studied the limit of large values of the compactification,
$L/\eta \gg 1$, where the leading term is given by Eq. \eqref{FCClargeL}.
Figure \ref{fig02} shows the behavior of the FC induced by the compactification as a function of $L/\eta$
and the axial magnetic flux. We have shown that both induced parts of the FC vanish for a massless fermion field.

Another important quantity that characterizes the fermionic vacuum is the VEV of the energy-momentum tensor.
The steps of the evaluation of each component of this tensor has been presented in Sec. \ref{EMTsec}
while in Sec. \ref{EMTsec2} we have presented combined expressions for both cosmic string in dS spacetime without compactification and topological
contribution of the energy-momentum tensor. The VEV of the energy-momentum tensor in dS space in the presence of a cosmic
string is given in a compact form \eqref{TS1} which includes the energy density, the radial and the
axial stresses. This contribution is an even function of the azimuthal magnetic flux. Some special cases of this expression have been evaluated.
For a massless fermionic field we have the expression \eqref{TS1massless}. Considering $r/\eta\ll 1$,
there is a divergence with a fourth power of the proper distance, \eqref{TS1R0}. On the other hand, for large
distances from the string, this contribution goes to zero as one can see in the Eq. \eqref{TS1largeR}. In Fig.
\ref{fig03} we have plotted the effect of the parameter $a_0$ on the energy-density
as a function of $r^2/\eta^2$ and $m\alpha$. For the contribution induced by the compactification,
the energy density and the radial stress are written in the compact form \eqref{TC1}. They are even functions of the magnetic fluxes in azimuthal and axial directions. For regions near the string, this
part of the energy-momentum tensor is given by \eqref{TC1R0} being divergent on the string if $|a_0|>(1-1/q)/2$
and finite if $|a_0|\leq(1-1/q)/2$. We have also decomposed this part of the energy-momentum tensor
into two parts: the first one is purely topological, \eqref{TC1a}, independent of the planar angle deficit, the
azimuthal magnetic flux and the radial coordinate, while the second one, \eqref{TC2},  is dependent on the compactification
and other parameters that characterize the system under consideration.
In the regime where $L/\eta \gg 1$, the latter has been given by
\eqref{TC2largeL}. In Fig. \ref{fig04} we have plotted the behavior of the energy density as a function of $L/\eta$ and $\tilde{\beta}$.

The azimuthal and axial stresses induced by the compactification have been evaluated in the subsections \ref{T22sec}
and \ref{T33sec}, respectively. An expression for the total azimuthal stress and the axial stress induced by the compactification has been written in compact forms
given by \eqref{T22total}  and \eqref{T33C}, respectively.
Some particular cases of these components of the energy-momentum tensor are presented in the Sec. \ref{EMTsec2}. 
Moreover, for a massless spinor field in the absence of the azimuthal flux, $a_0=0$, the azimuthal stress is
given by \eqref{T22massless} and \eqref{T22a0}, respectively.
The axial stress induced by the compactification considering a massless fermionic field
is given by \eqref{T33Cmassless}. This induced part of the axial stress has been decomposed into
two parts: a pure topological part, \eqref{T33C0}, independent of the the parameters
$q$, $a_0$ and the radial coordinate, and a part dependent on these parameters, \eqref{T33Cqa0}, where the latter is an even function of both magnetic fluxes. 

We have also verified that the components of the energy-momentum tensor obey the trace relation
\eqref{tracerelation} and the energy-momentum conservation in the covariant form. In particular, for a massless fermion field the energy-momentum tensor is traceless.

\appendix

\section{Relevant relations for Bessel, Hankel and Macdonald functions}
\label{ap2}
In this paper we use the following relations \cite{abramowitz,watson1995treatise,gradshteyn2000table,prudnikov2}
\begin{align}
H_\nu^{(2)}(x)&=\frac{2i}{\pi}e^{i\pi\nu/2}K_\nu(ix) \nonumber\\
|H^{(2)}_{1/2-im\alpha}(\gamma\eta)|^2&=\frac{4e^{\pi m\alpha}}{\pi^2}|K_{1/2-im\alpha}(i\gamma\eta)|^2 \nonumber\\
|H^{(2)}_{-1/2-im\alpha}(\gamma\eta)|^2&=\frac{4e^{\pi m\alpha}}{\pi^2}|K_{1/2+im\alpha}(i\gamma\eta)|^2\nonumber\\
K_\nu(x)&=K_{-\nu}(x)
\label{App}
\end{align}
\begin{align}
|K_{1/2-im\alpha}(ix)|^2-|K_{1/2+im\alpha}(ix)|^2&=
-i\left(\partial_x+\frac{1-2im\alpha}{x}\right)K_{1/2-im\alpha}(ix)K_{1/2-im\alpha}(-ix)
\label{App2}
\end{align}
\begin{align}
K_\nu(ix)K_\nu(-ix)=\int^\infty_0 du u^{-1}\int_0^\infty dy\cosh(2\nu y){\rm{exp}}\left[
{-2ux^2\sinh^2y-1/(2u)}\right]
\label{MacRep}
\end{align}
\begin{equation}
\left(u\partial_u+2\nu\right)e^{u^2}K_\nu(u^2)=2u^2e^{u^2}[K_\nu(u^2)-K_{\nu-1}(u^2)]
\label{Macder}
\end{equation}
\begin{equation}
\int_0^\infty dx \, x^{\alpha-1}e^{-px}K_\nu (cx)=\frac{\Gamma(\alpha-\nu) \, e^{-i\pi\nu}}{(p^2+c^2)^{\alpha/2}} \,
Q_{\alpha-1}^\nu\left(\frac{p}{\sqrt{p^2-c^2}}\right)
\label{leg1}
\end{equation}
\begin{equation}
Q_\nu^\mu(z)=\frac{e^{i\mu\pi}\sqrt{\pi} \, \Gamma(\nu+\mu+1)(z^2-1)^{\mu/2}}
{2^{\nu+1}\Gamma(\nu+3/2)z^{\nu + \mu+1}} \, F\left(1+\frac{\nu+\mu}{2},\frac{\nu +\mu+1}{2};\nu+\frac{3}{2};z^{-2}\right)
\label{leg2}
\end{equation}
\begin{align}
F(a,b;c;z)&=\frac{\Gamma(c)\Gamma(c-a-b)}{\Gamma(c-a)\Gamma(c-b)}\, F(a,b;a+b-c;1-z)\nonumber\\
&+\frac{\Gamma(c)\Gamma(a+b-c)}{(1-z)^{a+b-c}\Gamma(a)\Gamma(b)}\, F(c-a,c-b;c-a-b+1;1-z)
\label{hyper1}
\end{align}
\begin{equation}\label{Mac1}
K_\nu(e^{im\pi}x)=e^{-im\nu\pi}K_\nu(x)-i\pi\frac{\sin(m\nu\pi)}{\sin(\nu\pi)}
I_\nu(x)
\end{equation}
\begin{equation}
K_\nu(e^{i\pi} x)-K_\nu(e^{-i\pi}x)=-i\pi[I_\nu(x)+I_{-\nu}(x)]
\label{Mac2}
\end{equation}
\begin{equation}
\int^\infty_0 dp \, p \, e^{-a p^2}J_\nu ^2(p r)= \frac{e^{-r^2/2a}}{2a}I_\nu(r^2/2a)
\label{bess}
\end{equation}
\begin{align}
\epsilon_j [J_{\beta_1}(x)J'_{\beta_2}(x)-J'_{\beta_1}(x)J_{\beta_2}(x)]&=J^2_{\beta_1}(x)+
J^2_{\beta_2}(x)-\frac{2\beta_1+\epsilon_j}{x}J_{\beta_1}(x)J_{\beta_2}(x)
\label{bess-rec}
\end{align}
\begin{align}
J_{\beta_1}(pr)J_{\beta_2}(pr)&=\frac{1}{2p}\left(\frac{2\beta_1}{r}-\epsilon_j\partial_r\right)J^2_{\beta1}(pr)\nonumber\\
(\epsilon_j\beta_1-r\partial_r/2)e^{-r^2 \tau}I_{\beta_1}(r^2\tau)&=ze^{-z}[I_{\beta_1}(z)-I_{\beta_2}(z)]_{z=r^2\tau}\nonumber\\
\int_0^\infty dp \ p^2 J_{\beta_1}(pr)J_{\beta_2}(pr)e^{-2u\eta^2 p^2\sinh^2y}&=\frac{2\epsilon_j}{r}xe^{-x}
[I_{\beta_1}(x)-I_{\beta_2}(x)]
\label{bess-mod}
\end{align}
\begin{align}
(2\epsilon_j\beta_1+1)[I_{\beta_1}(x)-I_{\beta_2}(x)]&=2(x\partial_x-x+1/2)[I_{\beta_1}(x)+I_{\beta_2}(x)]
\label{rep-mod2}
\end{align}
\begin{equation}
\epsilon_j J_{\beta_1}(x)J_{\beta_2}(x)=\frac{1}{x}\left(\epsilon_j\beta_1-\frac{x}{2}\partial_x\right)J^2_{\beta_1}(pr).
\label{bess-2}
\end{equation}
\begin{equation}
\left(\frac{\epsilon_j\beta_1}{x}-\partial_x\right)e^{-x}I_{\beta_1}(x)=e^{-x}
[I_{\beta_1}(x)-I_{\beta_x}(x)]
\label{rep-mod}
\end{equation}

\section*{Acknowledgments}
E.A.F.B and A.M. would like to thank the Brazilian agency CAPES for financial support. A.M. also thanks financial support from the Brazilian agency CNPq and Universidade Federal de Pernambuco Edital Qualis A.


\end{document}